\documentclass[12pt,showpacs,twocolumn, preprintnumbers,aps,pre, reprint,superscriptaddress,nofootinbib]{revtex4-1}

\usepackage{graphicx}
\usepackage{subfigure}
\usepackage[labelfont={footnotesize,bf},textfont={footnotesize}]{caption}
\usepackage{color}
\usepackage{amsmath,amssymb,mathtools}
\usepackage{epstopdf}
\usepackage[linktocpage,colorlinks=true,linkcolor=blue,citecolor=blue, urlcolor = black, breaklinks=true]{hyperref}
\usepackage{url}
\usepackage{verbatim}
\usepackage{breakcites}
\usepackage[version=3]{mhchem}

\newcommand{\be}{\begin{equation}}
\newcommand{\ee}{\end{equation}}

\makeatother

\begin{document}

\preprint{}

\title{An intrinsic pink-noise multi-decadal global climate dynamics mode}

\author{Woosok Moon }
\affiliation{Department of Mathematics, Stockholm University 106 91 Stockholm, Sweden}
\affiliation{Nordita, Royal Institute of Technology and Stockholm University, SE-10691 Stockholm, Sweden}

\author{Sahil Agarwal}
\affiliation{Yale University, New Haven, USA}

\author{J. S. Wettlaufer}
\affiliation{Yale University, New Haven, USA}
\affiliation{Mathematical Institute, University of Oxford, Oxford, UK}
\affiliation{Nordita, Royal Institute of Technology and Stockholm University, SE-10691 Stockholm, Sweden}
\email[]{john.wettlaufer@yale.edu}

\date{\today}

\begin{abstract}
Understanding multi-decadal variability is an essential goal of climate dynamics. For example, the recent phenomenon referred to as the ``global warming hiatus" may reflect a coupling to an intrinsic, pre-industrial, multi-decadal variability process. Here, using a multi-fractal time series method, we demonstrate that forty-two data sets of seventy-nine proxies with global coverage exhibit pink noise characteristics on multi-decadal time scales.  To quantify the persistence of this behavior, we examine high-resolution ice core and speleothem data to find 
pink noise in both pre- and post-industrial periods. 
We examine the spatial structure with Empirical Orthogonal Function (EOF) analysis of the monthly-averaged surface temperature from 1901 to 2012.  The first mode clearly shows the distribution of ocean heat flux sinks located in the eastern Pacific and the Southern Ocean, and has 
pink noise characteristics on a multi-decadal time-scale. 
We hypothesize that this pink noise multi-decadal spatial mode may resonate with externally-driven greenhouse gas forcing, driving large-scale climate processes.
\end{abstract}

\pacs{}

\maketitle

A central question in contemporary climate science concerns the relative roles of natural climate variability and anthropogenic forcing.  Indeed, understanding the detailed human effect on global temperature is highly complex due to the nonlinear interactions of anthropogenic forcing with natural climate variability on multiple timescales, many of which transcend a typical human lifetime. The recent phenomena referred to as the ``global warming hiatus" \cite{Kosaka:2013aa, Karl:2015aa, Trenberth:2013aa, Watanabe:2014aa, Meehl:2014aa} is a compelling example emphasizing the potential of such interactions. Understanding the coupling between natural multi-decadal climate variability and anthropogenic forcing is a fundamental aspect of climate dynamics.

Here we describe a framework for characterizing the dynamics of natural global climate variability on multi-decadal timescales \cite{Mantua:2002aa, Schlesinger:1994aa, Salinger:2001aa}. There are many challenges associated with direct investigations of the physical and statistical characteristics of global observations on multiple timescales. First, the strong seasonal variability in observations, such as surface air temperature, hinders the detection of long-term spatiotemporal correlations \cite{Barnett:1978aa, Ludescher:2011aa, Agarwal:2012aa}. Moreover, there is a substantial land-ocean contrast in seasonal variability, making it difficult to extract the influence of climate variability on global scales. Second, the maximum length of the available station-based observations is only approximately 100 years, which may be insufficient to statistically discern multi-decadal variability.  At the same time, nonlinear interactions between natural and anthropogenic contributions to the multi-decadal variability found in these observations cannot be trivially disentangled. To overcome these obstacles, we analyze the data in a manner that enables us to exclude the contributions of the strong seasonality in the station-based observations.  Thus, we detect global multi-decadal timescales corresponding to pink noise dynamics, defined as having a power spectrum $S(f) \propto f^{-\beta}$, with frequency $f$ and $\beta \approx 1$, also generally termed $1/f$ noise when  0 $ < \beta < 2$ \cite[e.g.,][]{Davidsen:2002, Kendal:2011im, Blender:2011, Niemann:2013qy, Herault:2015fj, BrownnuttRMP}.
Furthermore, we analyze high-resolution proxy data spanning at least several hundred years to detect the footprint of these dynamics and to differentiate between anthropogenic forcing and natural climate variability.

We study the statistical characteristics of the decadal and multi-decadal variability of Earth's climate by analyzing the Goddard Institute for Space Studies (GISS) monthly-averaged surface temperature data from 1901 to 2012 \cite{GISTEMP, Hansen:2010aa}, and proxy data, such as $\delta^{18}O$ and $\delta^{13}C$, from ice-cores and speleothems from forty-two paleoclimate datasets (see Table 1). To examine the temporal dynamics of the data, we use Multi-Fractal Temporally Weighted Detrended Fluctuation Analysis (MF-TW-DFA) \cite{Agarwal:2012aa, Zhou:2010aa}. This methodology captures the statistical dynamics (e.g., white noise, red noise, degree of correlation) on multiple time scales. 
The veracity of the approach has been demonstrated in various fields, such as the study of Arctic sea ice extent \cite{Agarwal:2012aa}, sea ice velocity fields \cite{Agarwal:2017aa}, and even in the detection of exoplanets, in all cases solely using the data with no a-priori modeling \cite{Agarwal:2017ab}. This approach produces a statistical measure called the fluctuation function, $F_q (s)$, each moment of which $q$, is assessed on multiple time scales $s$, as described in \cite{Agarwal:2012aa, Zhou:2010aa} and Appendix \ref{sec:app} in more detail.  For intuition, one can think of the expectation value of $F_q (s)$ as the weighted sum of the auto-correlation function \cite[e.g.,][]{Lovsletten}.  
The dominant time scales in a system are the those where $F_q (s)$ versus $s$ changes slope and the individual slopes are associated with the statistical dynamics of a system. 

First, we analyze the GISS dataset by employing a new stochastic dynamical method of time-series analysis that was shown to capture the seasonal variability in monthly-averaged temperature data from decadal to 133 years \cite{Moon:2017aa}. This method centers on a periodic non-autonomous stochastic model for the observed deviation in the surface heat flux, $x(t)$, given by $\dot{x} = a(t)x + N(t)\xi(t) + d(\tau)$, where $a(t)$ and $N(t)$ are periodic functions with annual periodicity, $\xi(t)$ is stochastic noise, and $d(\tau)$ represents decadal forcing. Thus, the first two terms in the model explain the seasonal variability and the last term $d(\tau)$ captures the trans-seasonal variability. The approach provides analytical expressions for $a(t)$, $N(t)$ and $d(\tau)$, and reproduces the observed monthly statistics (Fig. 1 of \cite{Moon:2017aa}.)

\begin{figure}
    	\centering
    	\includegraphics[trim = 60 5 35 95, clip, width = 0.45\textwidth]{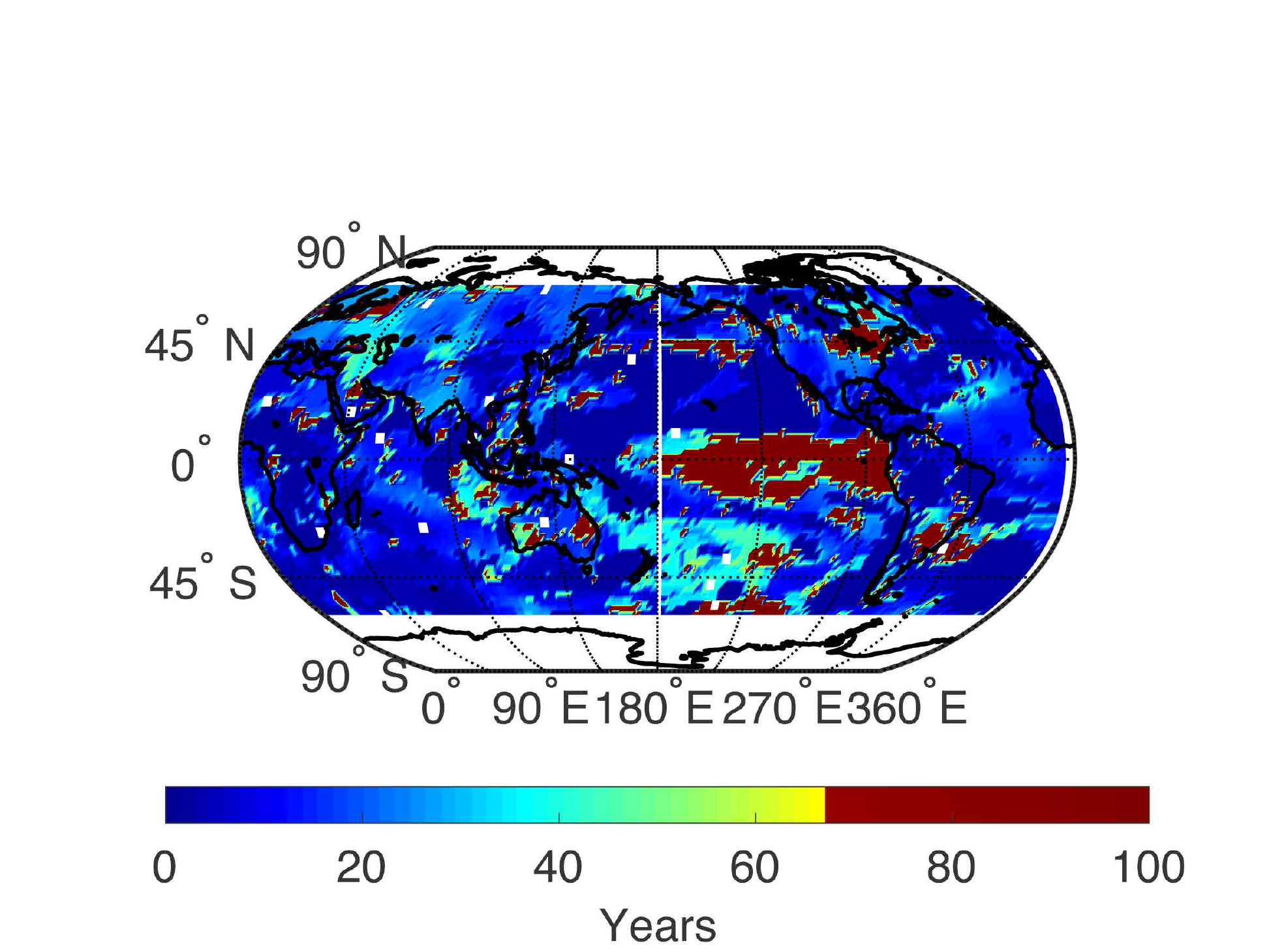}	
	\caption{Spatial distribution of the shortest timescale (in years) at which pink noise behavior appears in the GISS dataset. This transition takes place on multi-decadal timescales nearly everywhere. The red color denotes locations that do not show pink noise  characteristics on timescales up to 65 years (half of total length of the dataset), with the most prominent feature being in the tropical eastern Pacific. White regions show locations where continuous data are absent.}
\label{fig:globalmap}
\end{figure}

Second, we employ MF-TW-DFA to analyze the annual time-series for each latitude-longitude pair from the GISS dataset. A dominant signal at all locations is the presence of pink noise behavior ($\beta \approx 1$) on multi-decadal timescales. Pink noise, often referred to as ``ubiquitous noise'' \cite[e.g.,][]{Davidsen:2002, Kendal:2011im, Blender:2011, Niemann:2013qy, Herault:2015fj, BrownnuttRMP}, is observed in a wide range of systems, such as earthquakes, stellar luminosity, electronics, and climate on a variety of time scales \cite[e.g.,][]{Blender:2011}. We quantify the spatial structure of this statistical behavior by showing the timescales on a global map; Fig. \ref{fig:globalmap} shows the shortest timescale (in years) at which pink noise behavior appears in the data. Latitude-longitude pairs that do not show such behavior are shown in red, while points where no data was present are left blank.  Time scales greater than about 60 years are constrained by the finite length of the dataset. Thus, the colors on Fig. \ref{fig:globalmap} have two interpretations; pink noise from 1 to 60 years but no pink noise for longer times.  Because  both $d(\tau)$ and annual averaging of the data represent different forms of temporal filtering, they exhibit similar timescales for the global appearance of pink noise behavior, but we find quantitative but not qualitative differences (see Appendix \ref{sec:app}).
However, the value of using $d(\tau)$ is that it embodies the effects of seasonal stability and noise on annual and longer time scales. The point-wise values of $d(\tau)$ in the GISS dataset exhibit pink noise characteristics on decadal and multi-decadal timescales nearly everywhere on the globe. Dominant global climate variability phenomena such as the El Ni\~no-Southern Oscillation (ENSO) immediately emerge from this analysis. ENSO has been studied extensively and shown to influence global climate on time scales ranging from inter-annual to multi-decadal through atmospheric and oceanic teleconnections \cite[e.g.,][]{Liu:2007aa}. This phenomenon has also been related to global rainfall, a driver of global natural climate variability, which is a response to the regional amount of precipitation and evaporation, reflecting the variability in surface heat flux. 

\begin{figure}
    	\centering
    	\includegraphics[trim = 5 5 5 5, clip, width = 0.50\textwidth]{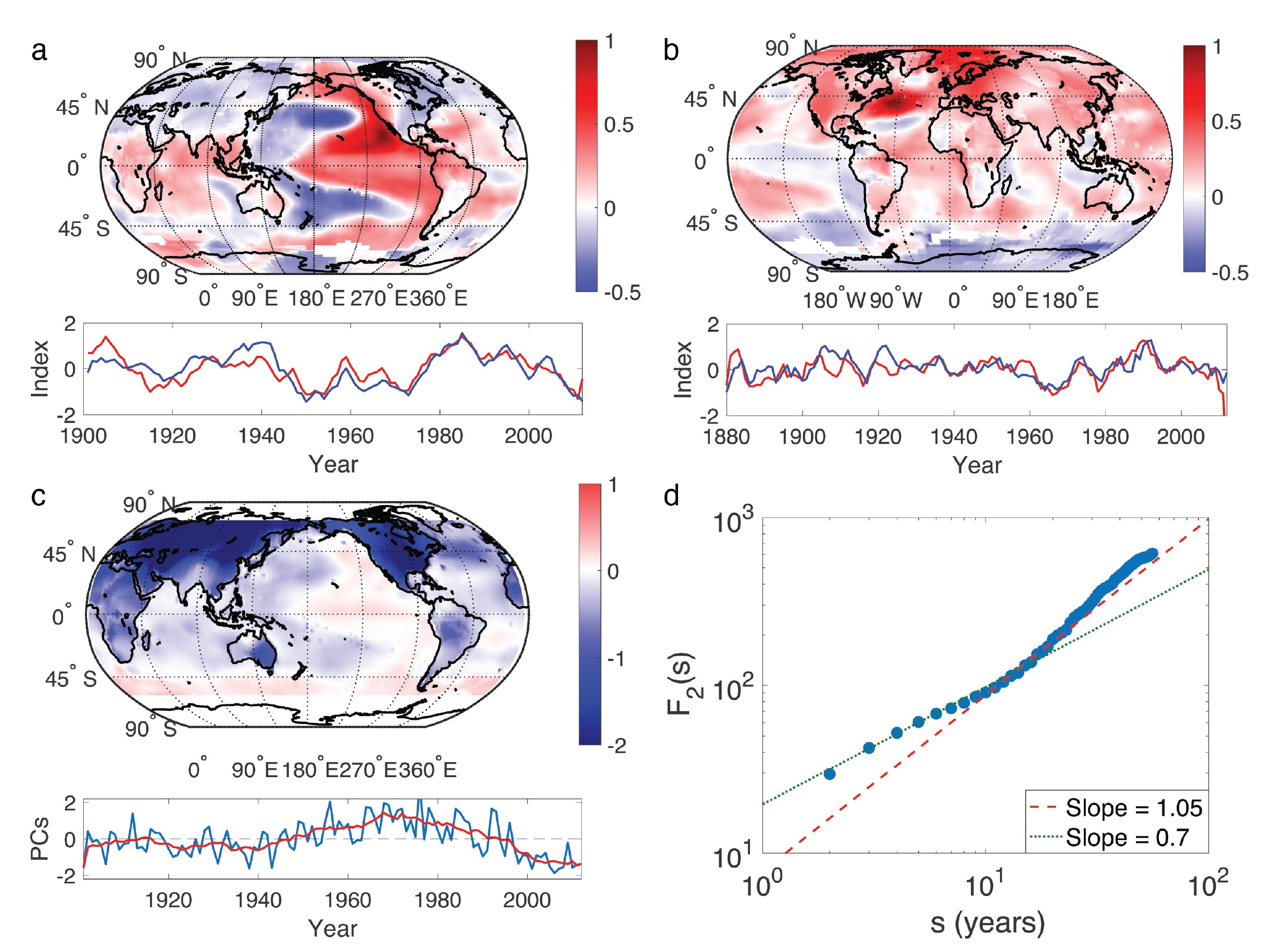}	
	\caption{ Spatial distribution of the values of $d(\tau)$ represented by one-point correlation maps and an Empirical Orthogonal Function (EOF) analysis using the Goddard Institute for Space Studies (GISS) surface monthly-averaged temperature from 1901 to 2012. Centered in the eastern Pacific at 120W, 20N, we calculate the correlation between the $d(\tau)$ at this position and that at any other position (a). The spatial distribution of the correlation is nearly identical to the dipole mode called the Pacific Decadal Oscillation (PDO) \cite{Kosaka:2013aa}. The newly constructed index (red), the normalized value of $d(\tau)|_{120W,20N} - d(\tau)|_{180E,40N}$, is compared with the traditional normalized PDO index (blue), which shows an excellent match. A similar one-point correlation map is constructed based on the geographic position at 50W, 38N and is shown in (b). This map is very similar to the SST pattern in the negative state of the NAO \cite{Hansen:2010aa}, as shown by the correlation between $d(\tau)|_{50W,38N} - d(\tau)|_{40W,50N}$ (red) and the normalized NAO index (blue). The EOF analysis is applied to the values of $d(\tau)$, with the leading mode explaining 21\% of the total variance, as shown in (c), along with the Principal Component (PC). This mode connects the major PDO region in the eastern Pacific to the Southern Ocean through a continuous same-sign region, as distinguished from the other areas. The time-series of the principle component of the mode is analyzed using MF-TW-DFA (d).  At lower frequencies the variability of $d(\tau)$ parallel's pink noise (red dashed line, $\beta=1$), with a crossover time of $\approx$ 15 years.}
	\label{fig:modes}
\end{figure}

To examine the spatial structure of $d(\tau)$, and whether it captures the principal contributions to decadal variability, we construct two one-point correlation maps. As seen in Fig. \ref{fig:modes}, $d(\tau)$ nearly mirrors two key decadal variability indices; (a) the Pacific Decadal Oscillation (PDO) \cite{Mantua:2002aa} and (b) the North Atlantic Oscillation (NAO) \cite{Wanner:2001aa}. We use Empirical Orthogonal Function (EOF) analysis \cite{Lorenz:1956aa} to determine the dominant spatial pattern. Fig. \ref{fig:modes}(c) shows the first EOF mode and explains 21\% of the total variance, with the rest of the modes characterized by shorter timescales (see Appendix \ref{sec:app}). This first mode connects the major PDO region in the eastern Pacific to the Southern Ocean region (also seen in simulations \cite[][]{Dijkstra}), and is very similar to the so-called ``hyper climate modes" \cite{Dommenget:2008aa} and the ``Inter-decadal Pacific Oscillation" (IPO) \cite{Salinger:2001aa} in the Pacific Ocean. The time-series of the Principal Component (PC) shows clear multi-decadal variability. We note that the sign of the mode changes from positive to negative at about the start of the new millennium. A negative sign denotes the intensification of the negative PDO in the North Pacific and the cooling of the Southern Ocean circumpolar region. Simulations \cite{Kosaka:2013aa, Trenberth:2013aa} have shown that the cooling of the eastern tropics is correlated with the ``global warming hiatus" and the average sea surface temperature trends from ten climate models, which capture the hiatus, are negative in the Eastern Pacific and Southern Ocean \cite{Watanabe:2014aa}. The leading EOF of $d(\tau)$ introduced here may be related to this hiatus. Fig. \ref{fig:modes}(d) shows the result of MF-TW-DFA using the time-series of the PC; the onset of pink noise behavior occurs after approximately 15-years, indicated by the fluctuation function mirroring the red dashed line denoting $\beta=1$. This noise behavior and its global presence on multi-decadal timescales raises the natural question; is pink noise dynamics an internal feature of the multi-decadal variability of our climate, or imprinted on the climate system by anthropogenic forcing?  We address this question by analyzing paleoclimate proxies.

Paleoclimate studies have been broadly successful in observing the long-term evolution and variability of Earth's climate  \cite[e.g.][]{Zachos:2001aa, Dansgaard:1993aa}. Due to their comparatively high resolution, we focus on the proxy data from speleothems and ice cores to (a) understand the observed pink signal in the GISS data, and (b) study the effect of anthropogenic climate change on natural climate variability. Our datasets cover a substantial swath of the globe; Asia, Europe, North America, Central America, South America, and Antarctica along with the Pacific Islands (see Appendix \ref{sec:app}). These data provide a long record of Earth's climate system, drawing from many sources dating back more than 100,000 years. Figure \ref{fig:paleo}(a) shows the fluctuation functions for the paleoclimate proxy data from various sources (see Appendix \ref{sec:app}). Two things are immediately evident: nearly all datasets show consistent pink noise behavior, and, as was observed in the GISS dataset, the transition timescale to this behavior depends on geographic location. 

To study the impact of anthropogenic forcing on internal climate variability and the observed pink noise behavior, we use MF-TW-DFA to analyze only data up to 1850 A.D.  Figure \ref{fig:paleo} shows that the fluctuation functions with and without the post-industrial period exhibit very little difference, indicating that the observed pink noise behavior is intrinsic to Earth's climate dynamics. In data from the last 80,000 years, we also find a timescale of approximately 1470 years (Fig. \ref{fig:DO}), the signal often ascribed to Dansgaard-Oeschger (DO) events \cite{Dansgaard:1993aa}.  We hypothesize the possibility of
 a stochastic resonance process due to the presence of pink noise on multi-decadal timescales as follows.  \citet{Nozaki:1998aa} showed that for noise with $1/f^{\beta}$, $0\le\beta\le2$, the noise intensity for which resonance takes place is {\it minimized} when $\beta \approx 1$ for relaxation oscillator dynamical systems, and DO events exhibit relaxation oscillation behavior \cite{Timmermann:2003aa, Boers:2017fi}.  Thus, the resonance efficiency is {\it maximal} for $\beta \approx 1$, and in all of these proxies DO events are preceded by pink noise on multi-decadal to centennial timescales, suggesting a much smaller pink noise intensity can lead to a new climatic state relative to other noise types, such as white noise.  Importantly, whether the DO events arise from stochastic resonance, a ``ghost-resonance'' or a related process is actively debated \cite[e.g.,][]{Wunsch, Ditlevsen:2005ct, Balenzuela:2012ya, Krumscheid:2015pr, Rypdal:2016bx, Boers:2017fi}, and here we emphasize that the time scale emerges from a stochastic data analysis method with no assumptions regarding periodicity.  We note further that in the dust flux data, which spans the last 800,000 years, we see a clear periodic 100,000-year signal related to the Milankovitch eccentricity cycle, providing a fidelity check for our methodology.

\begin{figure}
    	\centering
    	\includegraphics[trim = 10 10 35 60, clip, width = 0.45\textwidth]{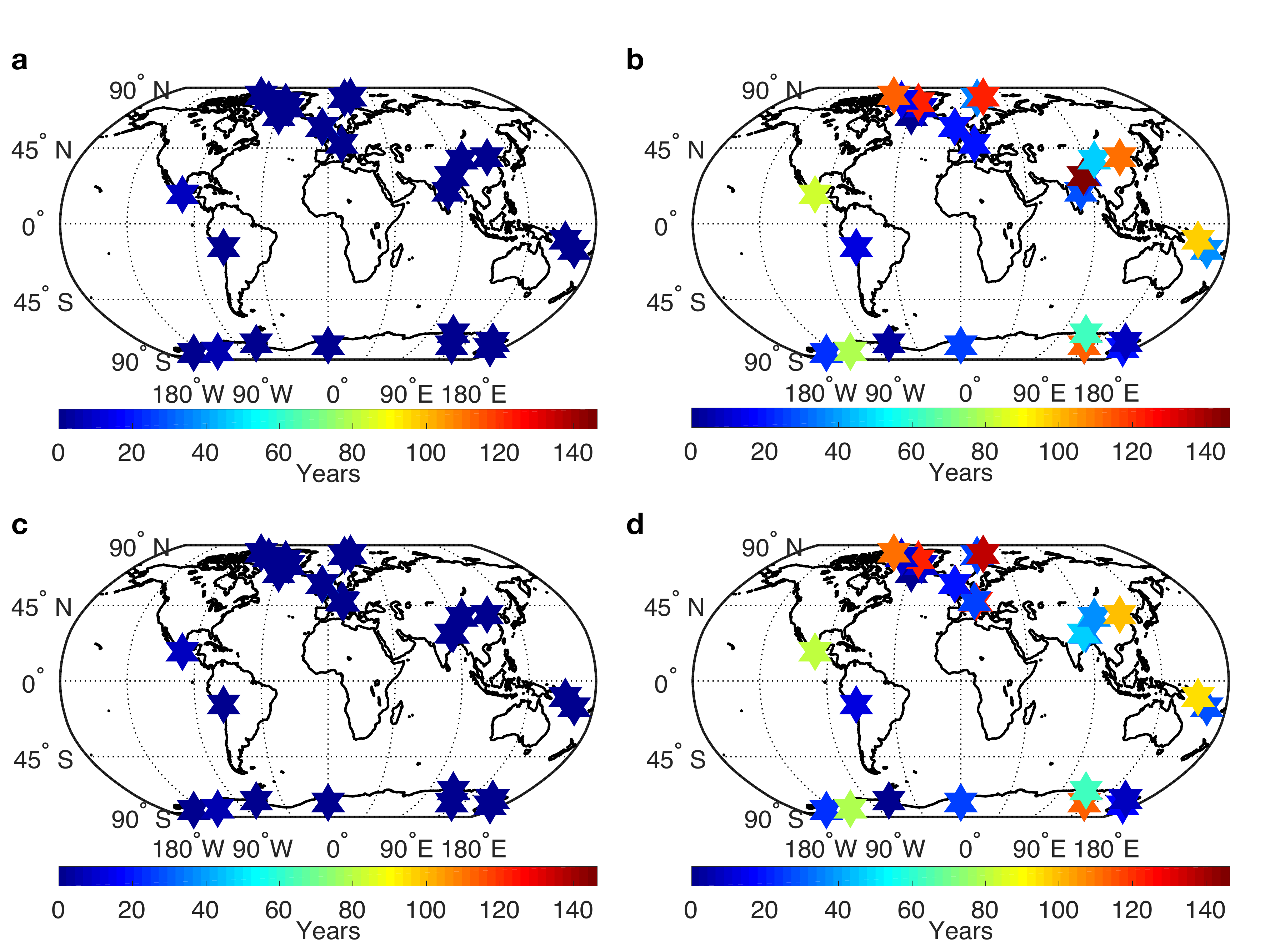}	
	\caption{The initial (a,c) and final (b,d) timescales exhibiting pink noise dynamics in the paleoclimate data across the globe (see Appendix \ref{sec:app}), where a \& b (c \& d) show the analysis for the complete dataset (after removing data from 1850-present), to enable us to distinguish between natural climate variability and anthropogenic forcing. There are no discernible differences between (a,b) and (c,d), implying that  pink noise dynamics are an internal characteristic of the Earth's climate system.}
	\label{fig:paleo}
\end{figure}

Proxies such as $\delta^{18}O$ and $\delta^{13}C$, from ice-cores and speleothems, are used to infer past temperature, amongst other climate variables.  Because temperature reflects heat flux at a given location, such flux dependent quantities are key mirrors of the climate system.  In the low (high) latitudes, heat fluxes drive precipitation-evaporation (freezing-melting).  Thus, global moisture fluxes are reflected with high fidelity in the ice core and speleothem proxy data and thereby encode aspects of climate variability.  For example, ENSO underlies major global rainfall patterns through atmospheric and oceanic teleconnections. Importantly, there are regional differences in the timescales over which the various paleoclimate proxies exhibit pink noise. Each precipitation-based proxy depends on the net heat flux at a given location and hence we expect regional variability of the pink noise timescales. 

\begin{figure}
    	\centering
    	\includegraphics[trim = 0 0 15 10, clip, width = 0.45\textwidth]{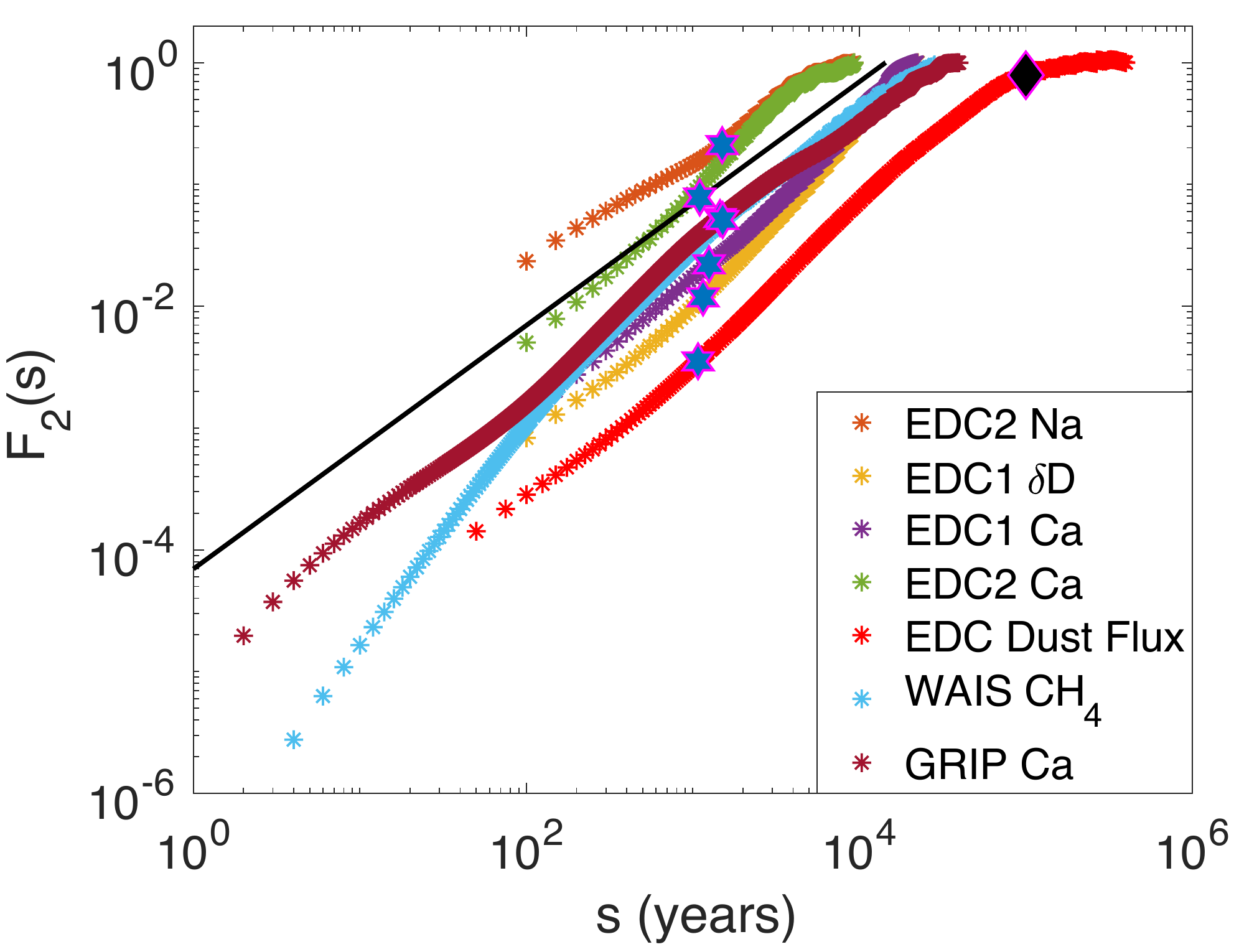}	
	\caption{Fluctuation functions $F_2(s)$ for paleoclimate proxy data (see inset and Table I in Appendix \ref{sec:app}) spanning at least the past 80,000 years. The red stars show the crossover timescale of $\sim$ 1470 years, associated with Dansgaard-Oeschger events.  For reference, the black diamond denotes the 100,000 year Milankovitch eccentricity cycle, the time scale of Late Pleistocene glaciations.  The black line has a slope of unity ($\beta =1$). }
	\label{fig:DO}
\end{figure}

Analysis of Sea Surface Temperature (SST) data has also revealed pink noise in the mid-latitudes \cite{Fraedrich:2003aa}, rationalized by a simple vertical diffusion model with a shallow mixed layer forced by random atmospheric motions \cite{Fraedrich:2004aa}. Essential here is the accumulation of the response from random atmospheric forcing due to the large heat capacity of ocean. This local variability in the mid-latitude and tropical oceans is transferred to the global scale via atmospheric teleconnections and ocean waves  \cite{Liu:2007aa, Dong:2015aa}. Here, this is reflected in our first EOF mode with a time evolution that shows pink noise statistics on multi-decadal time-scales. Moreover, the IPO, which we have shown mirrors our first EOF mode, is strongly linked to global precipitation \cite{Dong:2015aa}, consistent with the relationship between the pink noise behavior found in the proxies that reflect precipitation and the EOF mode. 

\citet{Kendal:2011im} have shown that both pink noise and fluctuation scaling (wherein the variance of a sequence of observations $x$ is related to the mean by a power law; Var$(x) \propto {\bar{x}}^b$) imply each other and can be explained by a central limit-like convergence theorem that establishes which Tweedie exponential dispersion models act as foci for this convergence \cite{Jorgensen:1994}.  
The duality between fluctuation scaling and pink noise not only provides a universal treatment of the statistics of the global mode that emerges from this wide range of data we have studied, but a common understanding of their non-Gaussianity. 

We note that the intrinsic nature of both the first EOF mode and the pink noise behavior suggest the intriguing potential of a resonance with external low-frequency forcing, such as that associated with anthropogenic effects. Such a resonance may underlie processes associated with global warming hiatus, emphasizing the importance of understanding internal multi-decadal variability.

Finally, non-autonomous stochastic differential equations constitute a key organizing center of our approach \cite{Moon:2017aa}, and they are also central to the so-called supersymmetric theory of stochastics \cite[e.g.,][]{Ovchinnikov:2012nt, Ovchinnikov:2016}.  That approach argues that pink noise is a manifestation of the spontaneous breakdown of topological supersymmetry.  However, to ascribe the associated Goldstone modes to specific climate processes is too speculative at present, although the breaking of time-reversal symmetry by Earth's rotation has been shown to provide a topological origin for equatorially trapped waves \cite{Marston}.  Therefore, understanding the origin of the emergence the decadal modes in the climate system that we have observed here may be fruitfully pursued along these lines.  

\begin{acknowledgements}
WM acknowledges a Herchel-Smith postdoctoral fellowship for support. SA acknowledges David Crighton fellowship for support. SA and JSW acknowledge NASA Grant NNH13ZDA001N-CRYO.  JSW and WM acknowledge Swedish Research Council grant no. 638-2013-9243, and JSW a Royal Society Wolfson Research Merit Award for support. 
 \end{acknowledgements}
 
 \appendix

\section{\label{sec:app}}
 
 \subsection{GISS Dataset}
NASA Goddard's Global Surface Temperature Analysis (GISTEMP) provides an estimation of the anomalous change of global surface temperature, relative to the 1951-80 base time period mean, from 1880 to present. This data set has been constructed by collecting data from many national meteorological  services around world, including Global Historical Climatology Network (GHCN) Vs. 3, United States Historical Climatology Network (USHCN) data, and SCAR (Scientific Committee on Antarctic Research) from Antarctic stations. The spatial and temporal resolutions are 2 by 2 degree and a month, respectively. 

\subsection{Paleoclimate Proxy Data}
\begin{figure}
\centering
   \includegraphics[width = 0.45\textwidth] {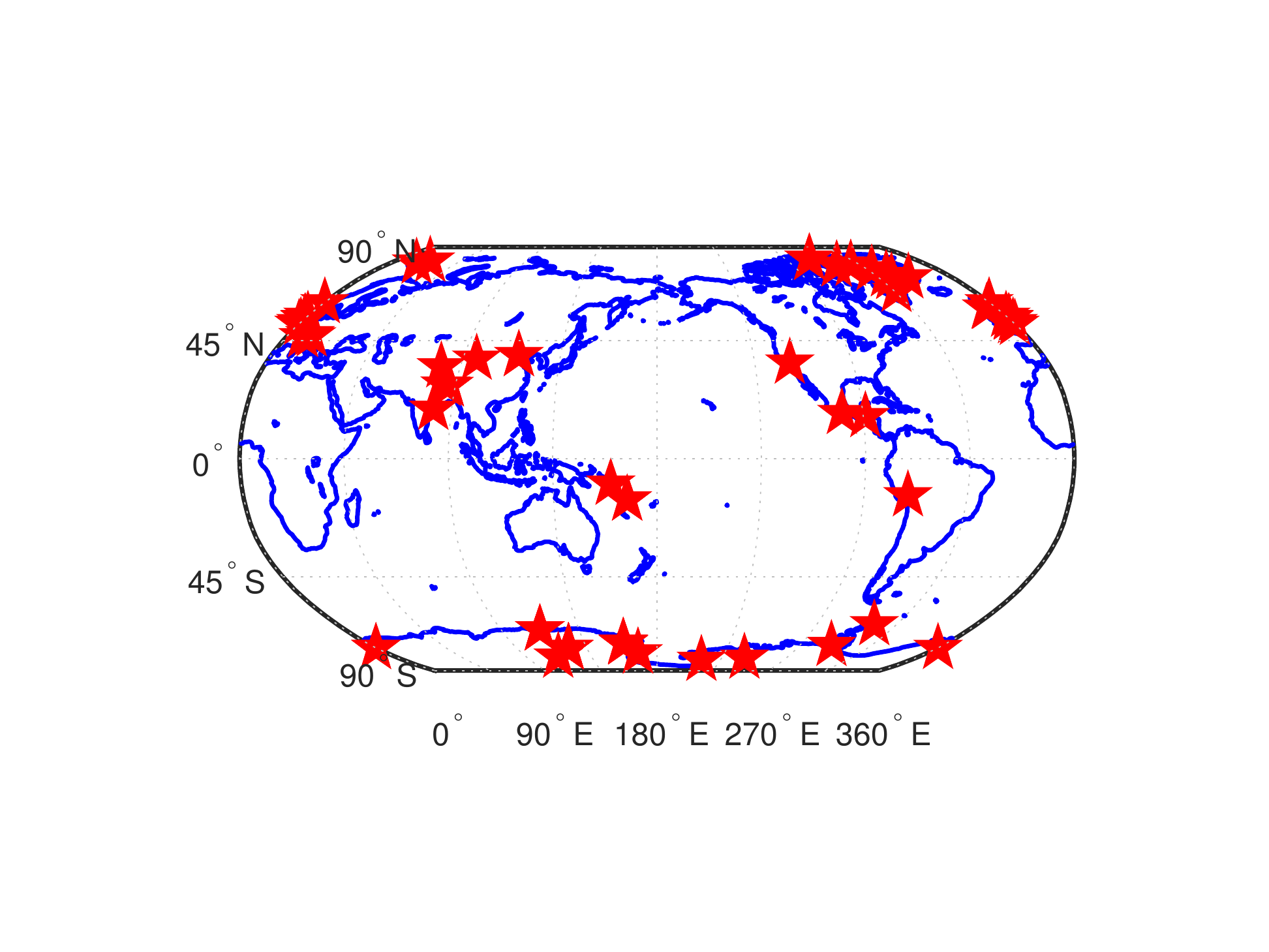}
\caption{The stars on the world map show the locations of all the ice-core/speleothem datasets used in this study.}
\label{fig:GeoPaleo}
\end{figure}

We used the speleothem and ice-core data from the National Center for Environmental Information at the National Oceanic and Atmospheric Administration. The datasets include isotope concentrations of $\delta^{18}$O, $\delta^{13}$C, $\delta$ D, Ca, Na, CO$_\text{2}$, SO, SO$_\text{4}$, NO, N$_\text{2}$O, CH$_\text{4}$, Cl, dust flux, electrical conductivity measurements (ECM), temperature, stalagmite growth-rate and mean annual precipitation. These isotopic concentrations along the depth of the ice-cores/speleothems are used as proxies for temperature in the Earth's past climate. Table \ref{paleotable} shows all the datasets used in this study, and  Figure \ref{fig:GeoPaleo} shows the locations of these data.

\subsection{Multi-Fractal Temporally Weighted Detrended Fluctuation Analysis}

Central aspects of any geophysical process include nonlinear behavior and the confluence of  multiple timescales with  long term trends. These multiple timescales are important to understand the plethora of other physical processes that give rise to the process in question. One would also like to understand the dynamics of the process on these multiple time scales. Perhaps the most commonly used method in this regard is the two-point autocorrelation function. Unfortunately, this method can be severely limited if there are any trends or seasonality present in the data. Moreover, perhaps the most serious limitation is that it assumes that there is only a single time scale that characterizes the data. To overcome these drawbacks, we use the multifractal temporally weighted detrended fluctuation analysis. This methodology exploits the intuition that points closer in time are more related to each other than points farther in time. Moreover, it does not limit the number of timescales that may be present in the system under study. The method has four major steps:

\begin{enumerate}
\item Construct a non-stationary profile $Y(i)$ as cumulative sum of the time series $X_i$,
\begin{equation}
Y(i)= \sum_{k = 1}^{i}(X_k - \bar{X}), \qquad k=1,2,...,N
\end{equation}

\item Divide this profile into $N_s= int(N/s)$ non-overlapping segments of equal length $s$, where $s$ is an integer and $s\times \Delta t$ represents a timescale with $\Delta t$ being the time resolution of the time series. Since the length of the time series may not be an exact multiple of $s$, this procedure is repeated by starting from the end of the profile and returning to the beginning and hence creating $2N_s$ segments.

\item A point-by-point approximation of the profile $\hat{y}(i)$ is made of the original profile using a moving window of length $s$, with weights given by the distance between points $i$ and $j$, such that $|i-j| \le s$. A larger (smaller) weight $w_{ij}$ is given to $\hat{y}(i)$ if $|i-j|$ is small (large). This approximated profile is then used to compute the variance of the original profile versus the approximated profile spanning up $(\nu = 1,2,...,N_s)$ and down $(\nu = N_s+1,N_s+2,...,2N_s)$ as follows

\begin{eqnarray}
\textnormal{Var}(\nu, s) \equiv & \frac{1}{s}  \sum_{i=1}^{s} \{ Y([\nu - 1]s + i) - {\hat{y}}([\nu-1]s +i) \}^2 \nonumber \\
&\textnormal{for $\nu = 1,...,N_s$, and} \nonumber \\
\textnormal{Var}(\nu, s)  \equiv & \frac{1}{s}  \sum_{i=1}^{s} \{ Y(N-[\nu - N_s]s + i) - {\hat{y}}(N-[\nu-N_s]s +i)\}^2\nonumber \\
&\textnormal{for $\nu =  N_s + 1,...,2 N_s$.}
\label{eq:varTW}
\end{eqnarray}

\item This allows us to obtain a generalized fluctuation function as

\begin{equation}
F_q (s) \equiv \left[ \frac{1}{2 N_s} \sum_{\nu=1}^{2 N_s} \{ \textnormal{Var}(\nu, s)\}^{q/2} \right]^{1/q}, 
\label{eq:fluct}
\end{equation}
\end{enumerate}
the behavior of which depends on window length $s$, as well as the moment $q$. The principal focus is to study the scaling of $F_q(s)$,  which is characterized by a generalized Hurst exponent $h(q)$ through 
\begin{equation}
F_q (s) \propto s^{h(q)} .  
\label{eq:power}
\end{equation}
For a mono-fractal time series, $h(q)$ is independent of the moment $q$. In this case, $h(q)$ is equivalent to the classical Hurst exponent $H$. One generally studies the second moment of the fluctuation function as $h(2)$ can be related to the decay of the power spectrum $S$. If $S(f) \propto f^{-\beta}$, where $S(f)$ is the spectral density at frequency $f$, then $h(2) = (1 + \beta)/2$. For a white noise process, $\beta = 0$ and hence $h(2) = 1/2$. Similarly, for a red noise process $\beta = 2$ and therefore $h(2) = 3/2$. The data is said to be long term correlated for $0.5<h(2)<1$ and short term correlated for $h(2) > 1$. The dominant time scales in the data are where the fluctuation function $\log_{10}F_2(s)$ changes slope with respect to $\log_{10}s$. A crossover in the slope of a fluctuation function is calculated if the change in slope of the curve exceeds a threshold of $C_{th} = 0.02$. The range of time scales showing pink noise behavior is defined by the slope of the fluctuation function residing between $0.8 \le \beta \le 1.2$.

Here and in the main paper we show only the second moment of the fluctuation function $F_2$.   Our main motivation in showing only the second moment is its direct connection to the power spectrum, which is more familiar to the general reader. However, although we only show $F_2$, for all datasets we checked $h(q) \textnormal{~vs~} q$, i.e., the variation of Generalized Hurst exponents with the moments of the fluctuation functions to ensure that multifractal behavior exists.

\begin{figure}
\centering
   \includegraphics[width = 0.45\textwidth] {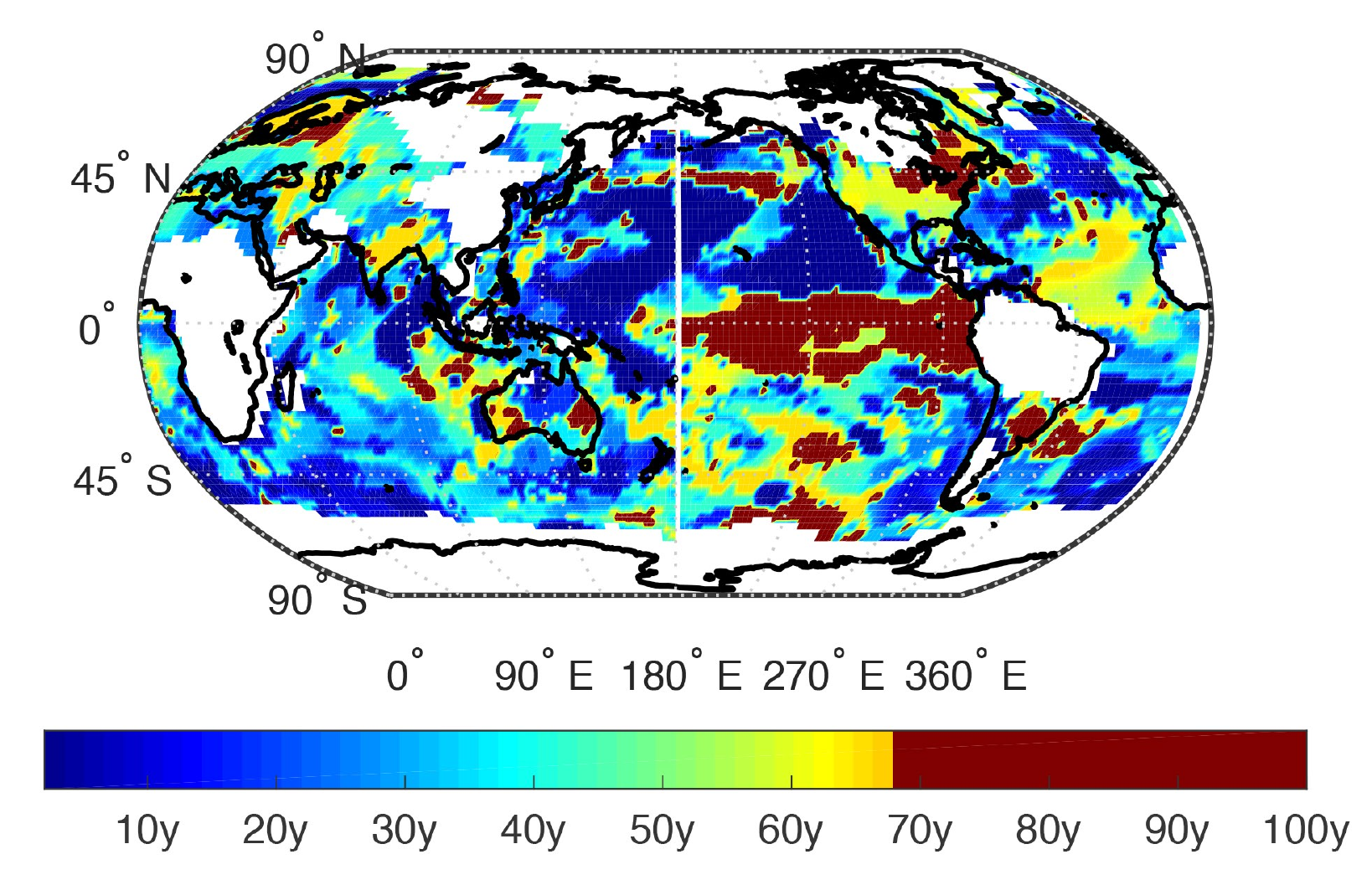}
\caption{Spatial distribution of the longest timescale up to 66 years (in years, starting from the shortest timescale from Fig. 1 in main text) over which the GISS dataset exhibits pink	 noise characteristics. This spatial distribution confirms the presence of pink noise characteristics on multi-decadal timescales. The colors follow the same scale as in Fig. 1 of the main text. }
\label{fig:GISS_TS}
\end{figure}

In a previous study \cite{Agarwal:2012aa}, we showed the effect of the presence of a strong seasonal cycle in the data on fluctuation functions. Due to the strength of the seasonal cycle, the fluctuation function saturates after the seasonal cycle timescale, thereby, masking the longer timescales that may be present in the data. To make sure that the fluctuation functions exhibiting pink noise behavior in the monthly dataset for $d(\tau)$ are not affected by seasonality of the data, we check for re-entrant behavior in the data.  Figure \ref{fig:GISS_TS} shows the spatial distribution over the globe of the longest timescale starting from the shortest timescale (from Fig. 1 in the main text), over which the GISS dataset exhibits pink noise characteristics. Taken together, these two figures (Fig. 1 and  Figure \ref{fig:GISS_TS})  show the multi-decadal global distribution of pink noise. 

In our analysis of sea ice data we demonstrated that the use of regular Detrended Fluctuation Analysis (DFA) can obscure the longer timescales due, among other things, to the rapidly varying fluctuation functions in that method \cite{Agarwal:2012aa}. Moreover, whereas regular DFA is useful only up to $N/4$, where $N$ is the length of the time-series, MF-TW-DFA is informative up to $N/2$. We showed that even ninth order DFA (DFA-9) was not able to successfully capture the long timescales in the Arctic sea ice extent data. In light of this, we note that  \citet{Huybers:2006aa} 
use DFA-1 (i.e. 1st order polynomials to fit the profile in this methodology) as a check on their results using Thomson's multitaper spectral method. Hence, given our finding of the lack of fidelity of higher order DFA, it may be of interest to revisit the intercomparison of the multitaper method with both (very) high order DFA and MF-TW-DFA given that the latter has been shown to resolve long timescales has much higher fidelity than the former.

\subsection{Statistical Significance of exponents $\beta$}

The robustness of our exponents is shown in Figures \ref{fig:knn1} - \ref{fig:knn5} below, not originally included in the SM, but now added.  We perform a series of calculations with the GISS dataset and use the Central Limit Theorem to show that $1/f$ noise is robust on the timescales shown in Figure 1 of the main text.   
The procedure is as follows:

\begin{enumerate}

\item For a region on the globe, define an array (lat-lon grid) with a synoptic scale with $6^{\circ} \times 6^{\circ}$ resolution. The original resolution of the GISS dataset is  $2^{\circ} \times 2^{\circ}$.
\item For a point in this array, compute its $k$-nearest neighbors ($\equiv$ KNN).
\item Compute the mean of the slope or exponent of these KNN.
\item Repeat steps 2 and 3 for each point in the array.
\item Figures \ref{fig:knn1} - \ref{fig:knn5} show this for 3 values of $k$; 10, 30 and 50, for 5 regions -- the complete globe and four quarters.  
\end{enumerate}

\begin{figure*}[htbp]
\centering
   \includegraphics[width = 0.8\textwidth]{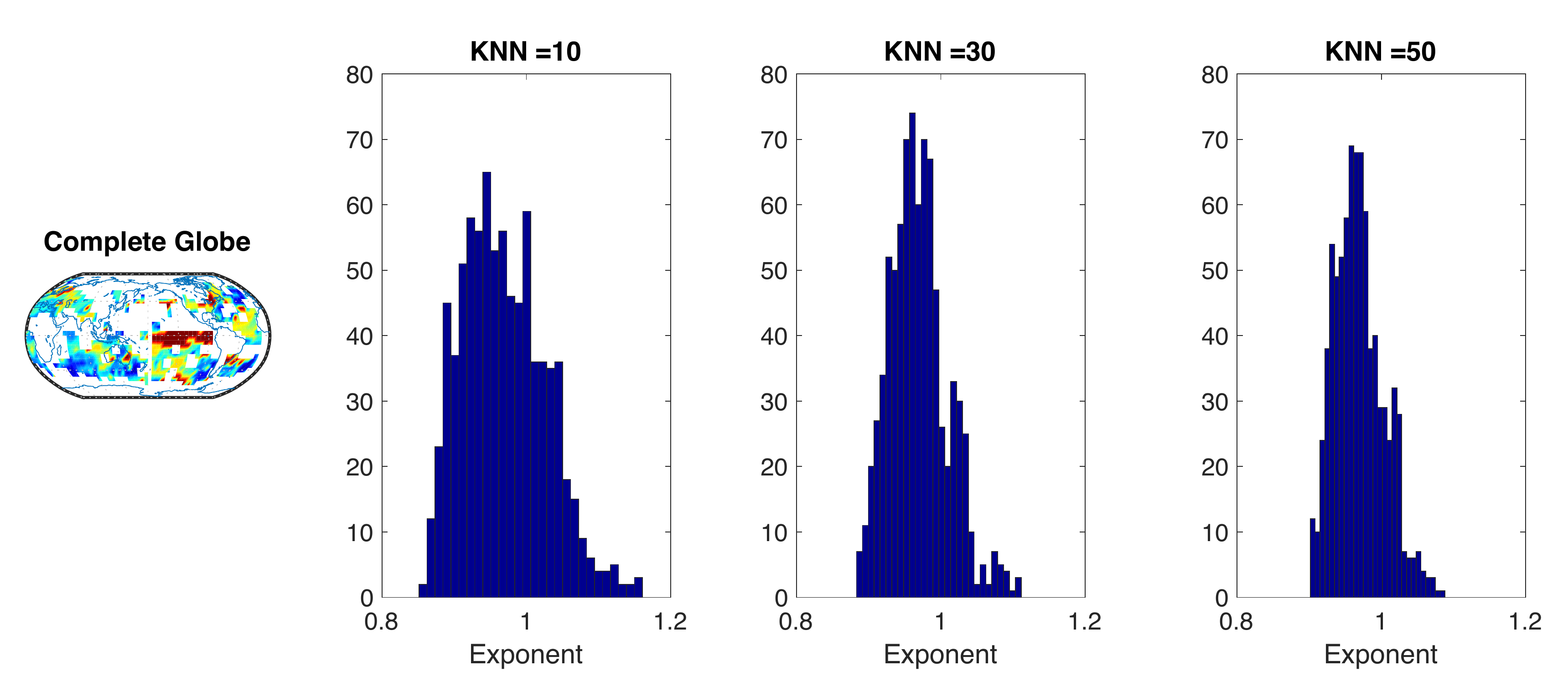}
\caption{The map on the left shows the region under consideration to test the robustness of the determination of $1/f$ noise. The three histograms show the distribution of mean exponents $\beta$ of nearest neighbors to each point in the grid for three values of $k$. As the number of nearest neighbors increases, the distribution of exponents becomes more sharply peaked around {\bf 1} thus showing that $1/f$ noise is robust on these timescales.}
\label{fig:knn1}
\end{figure*}

\begin{figure*}[htbp]
\centering
   \includegraphics[width = 0.8\textwidth] {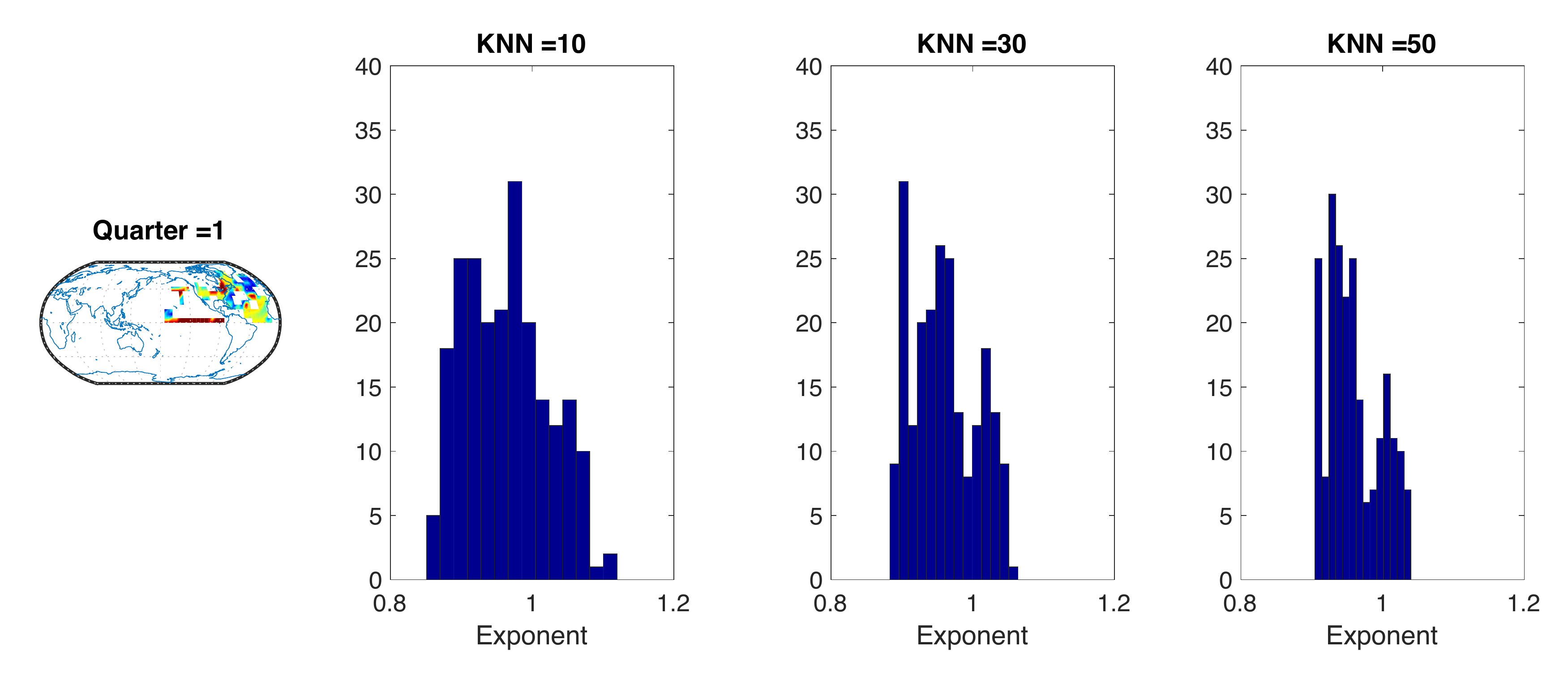}
\caption{Same as Figure \ref{fig:knn1}, but for the 1$^{\text{st}}$ quarter of the globe, defined as $0^\circ \le \text{Latitude} \le 90^\circ$, and $-180^\circ \le \text{Longitude} \le 0^\circ$. }
\label{fig:knn2}
\end{figure*}

\begin{figure*}[htbp]
\centering
   \includegraphics[width = 0.8\textwidth]{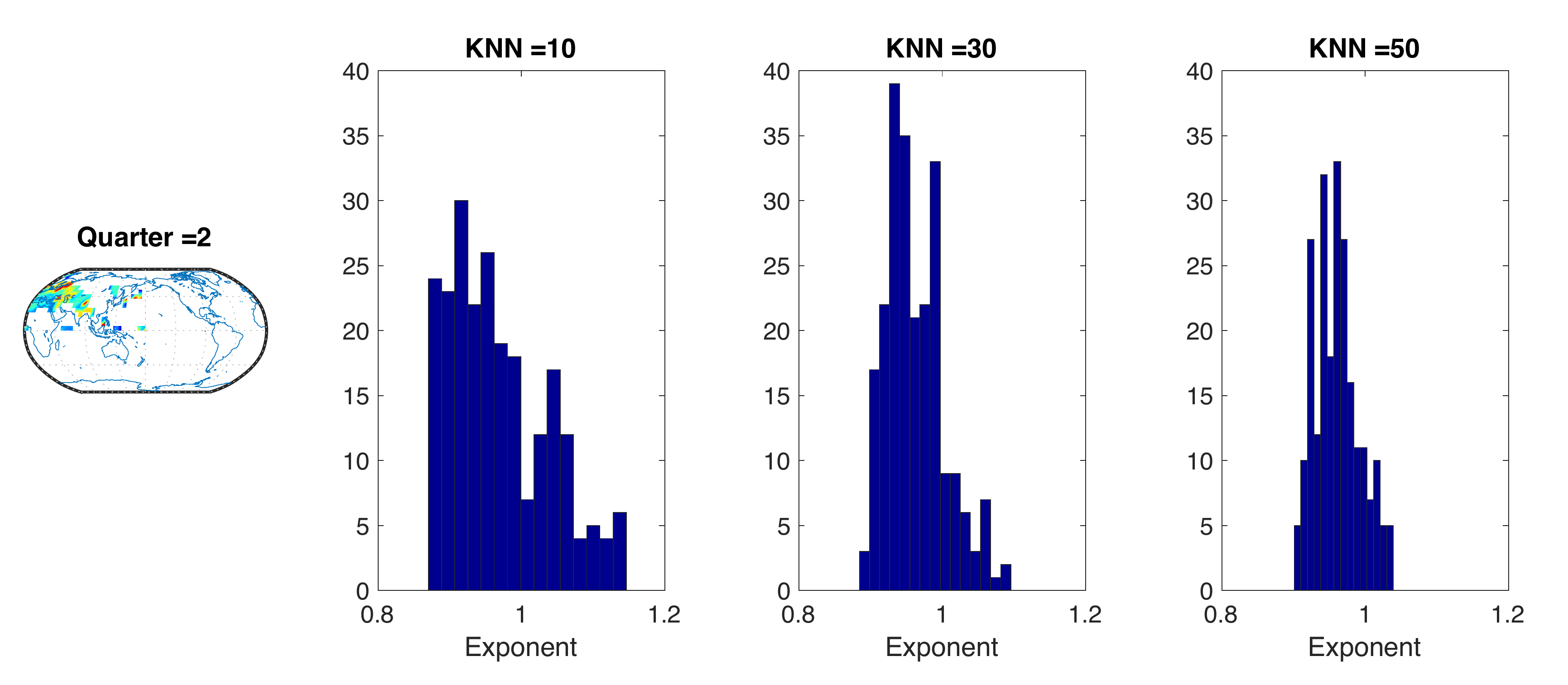}
\caption{Same as Figure \ref{fig:knn1}, but for the 2$^{\text{nd}}$ quarter of the globe, defined as $0^\circ \le \text{Latitude} \le 90^\circ$, and $0^\circ \le \text{Longitude} \le 180^\circ$.  }
\label{fig:knn3}
\end{figure*}

\begin{figure*}[htbp]
\centering
   \includegraphics[width = 0.8\textwidth]{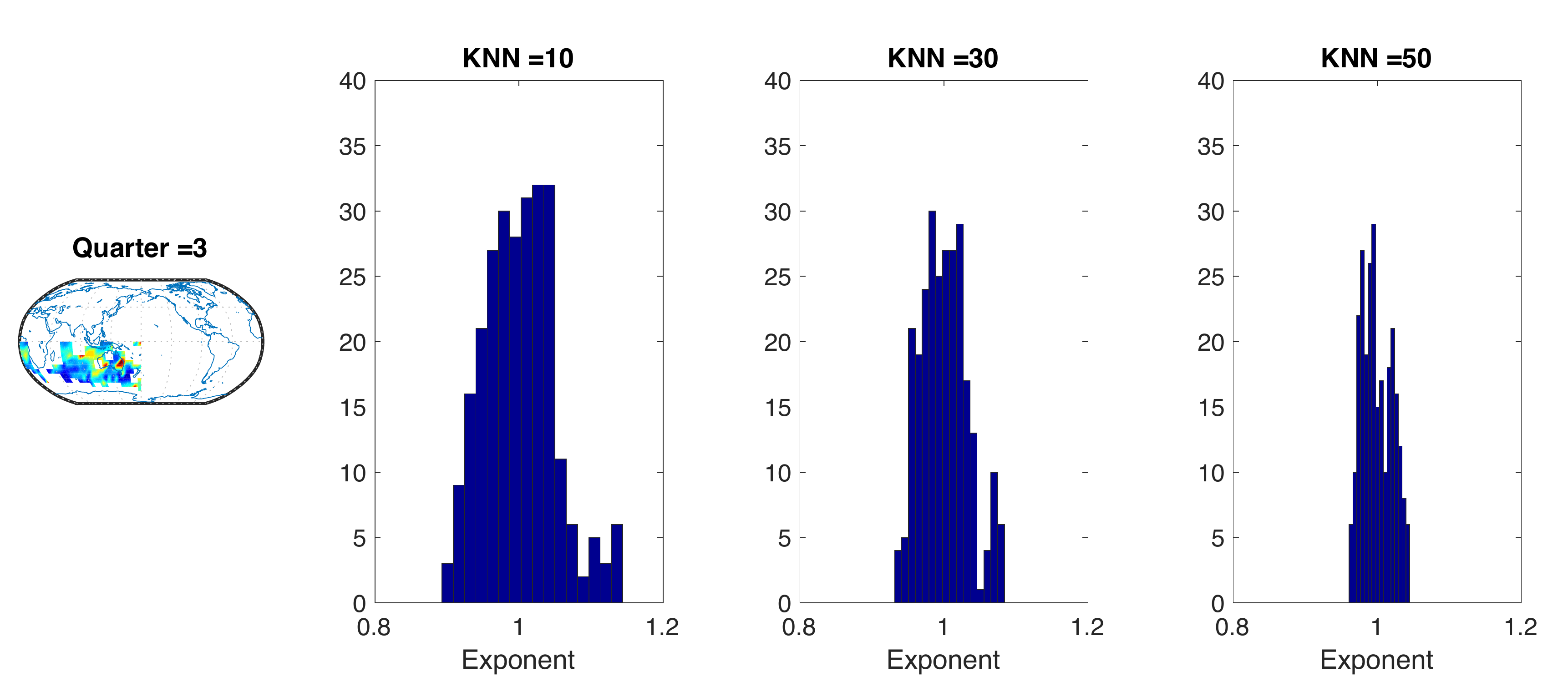}
\caption{Same as Figure \ref{fig:knn1}, but for the 3$^{\text{rd}}$ quarter of the globe, defined as $-90^\circ \le \text{Latitude} \le 0^\circ$, and $0^\circ \le \text{Longitude} \le 180^\circ$.  }
\label{fig:knn4}
\end{figure*}

\begin{figure*}[htbp]
\centering
   \includegraphics[width = 0.8\textwidth]{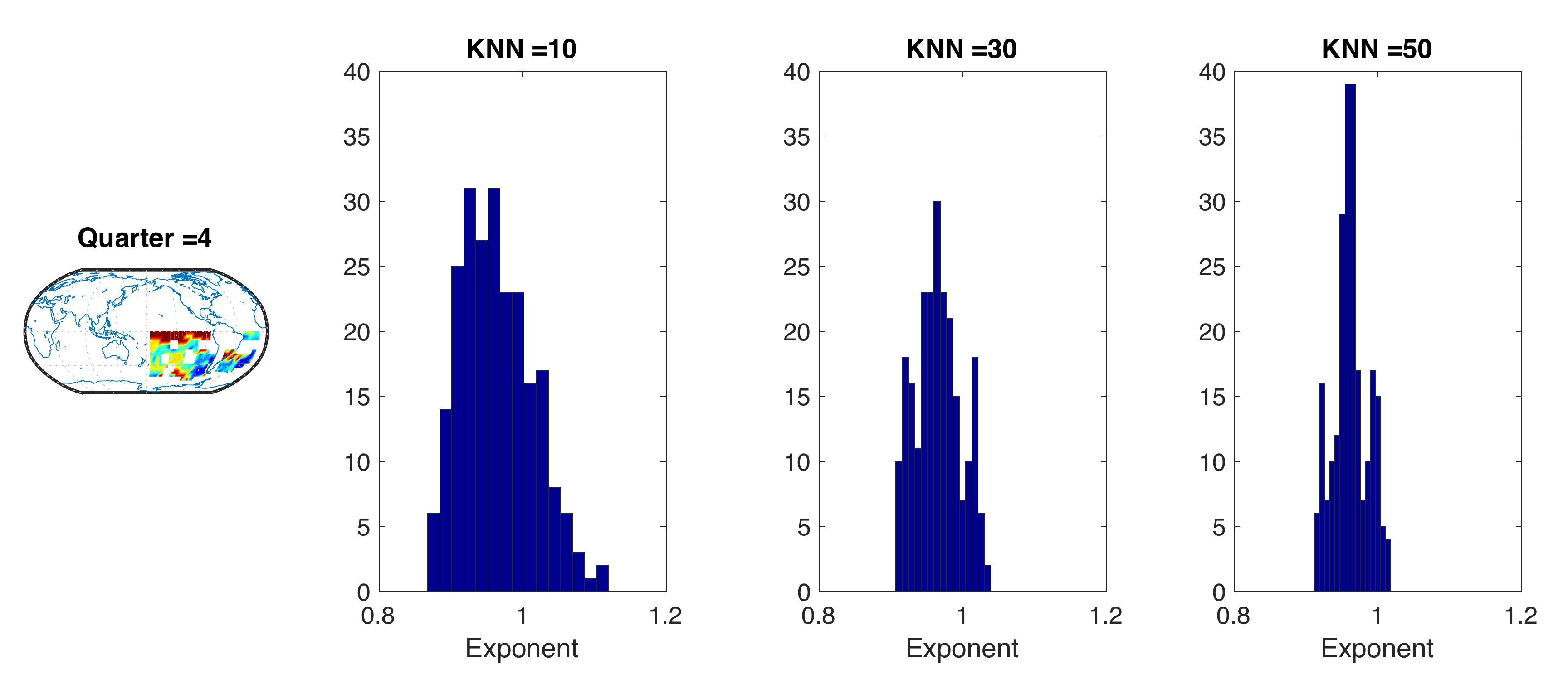}
\caption{Same as Figure \ref{fig:knn1}, but for the 4$^{\text{th}}$ quarter of the globe, defined as $-90^\circ \le \text{Latitude} \le 0^\circ$, and $-180^\circ \le \text{Longitude} \le 0^\circ$. }
\label{fig:knn5}
\end{figure*}

\subsection{Stochastic time-series analysis method}
This new stochastic time-series analysis can construct the monthly stability $a(t)$, the noise intensity $N(t)$, and the long-term 
forcing $d(\tau)$, from a monthly-averaged climate variable. Here we use monthly averaged anomalous surface air temperature $x^k_i$,
where $k$ and $i$ are the indices for month and year, respectively. The detailed theoretical background and implementation of the 
time-series method for seasonal variability are described in reference \cite{Moon:2017aa}.\\

\underline{\textit{The basic statistics of a time series $x^k_i$}} \\
The time-series method begins by calculating three statistical moments from monthly averaged data: 
({\em i}) monthly variances $S(k)$ from January ($k=1$) to December ($k=12$) ({\em ii}) the correlation between two adjacent months $A(k)$, and ({\em iii})
the one or two year autocorrelation $B(k)$; 
\begin{align}
 &\langle x^2(k) \rangle \simeq \frac{1}{M-1}\sum^{M}_{i=1}x^k_i x^k_i \equiv S(k), \nonumber \\
 &\langle x(k)x(k+1) \rangle \simeq \frac{1}{M-1}\sum^{M}_{i=1}x^k_i x^{k+1}_i  \equiv A(k) \qquad \textnormal{and} \nonumber \\
 &\langle x(k)x(k+12m) \rangle \simeq \frac{1}{M-m-1}\sum^{M-m}_{i=1} x^k_i x^k_{i+m} \equiv B(k), 
\end{align}
for the $k^{th}$ month, with $12m$ for $m$ years as one step is one month. 
Thus, for $B(k)$ we can choose $m$ to be 1 or 2 accordingly, and $S(k)$, $A(k)$ and $B(k)$ are periodic with period 1 year. 

\underline{\textit{Monthly stability $a(t)$}} \\
We introduce
\begin{align}
&G(t) = \frac{1}{\Delta t}\frac{S(t)-A(t)}{S(t)} \qquad \textnormal{and}\nonumber \\
&H(t)=\frac{S(t)-B(t)}{S(t)},
\end{align}
and then calculate the steady-state periodic solution $P(t)$ of the periodic non-autonomous ordinary differential equation, $dP/dt = -G(t)P(t)+H(t)$.
The monthly stability is then $a(t)=1/P(t)\times\left[H(t)-G(t)P(t)-1\right]$. \\

\underline{\textit{Noise intensity $N(t)$}} \\
Having constructed $a(t)$, we introduce new data $y(t)=x(t+\Delta t)-x(t)-a(t)x(t)\Delta t$ that can be 
represented as $N(t)\Delta W + d(\tau)\Delta t$. Here $\Delta W$ is a Brownian increment where $\langle \Delta W(t) \Delta W(t')\rangle = \delta (t-t')$,
giving
\begin{eqnarray}
 N^2(t)=\frac{1}{\Delta t}\left(\langle y^2(t)\rangle - \langle y(t)t(t+\Delta t) \rangle \right).
\end{eqnarray}

\underline{\textit{Long-term forcing $d(\tau)$}}\\
Having determined $a(t)$ and $N(t)$, we calculate $d(\tau)$ using $x(t+\Delta t)-x(t)-a(t)x(t)\Delta t-N(t)\Delta W$. 
Because $d(\tau)$ is decadal forcing, it will be approximately constant on annual time scales and hence we 
take the annual average of the remaining term, which leads to

\begin{eqnarray}
d(\tau) &=& \left \langle \frac{1}{n}\sum^{n}_{k=1}x(t_k+T)-\frac{1}{n}\sum^{n}_{k=1}x(t_k) \right. \nonumber \\
&& \left. -\frac{1}{n}\sum^{n}_{k=1}\int_{t_k}^{t_k+T}a(t)x(t)dt -\frac{1}{n}\sum^{n}_{k=1}\int_{t_k}^{t_k+T}N(t)dW \right \rangle \nonumber \\
&&\hspace{-10mm} = \frac{1}{n}\sum^{n}_{k=1}x(t_k+T)-\frac{1}{n}\sum^{n}_{k=1}x(t_k)-\frac{1}{n}\sum^{n}_{k=1}\int_{t_k}^{t_k+T}a(t)x(t)dt. \nonumber \\
 \label{eq:d}
\end{eqnarray}

More detailed derivations and examples are described in the supplementary material of \citet{Moon:2017aa}.

\begin{figure}
\centering
   \includegraphics[width = 0.45\textwidth]{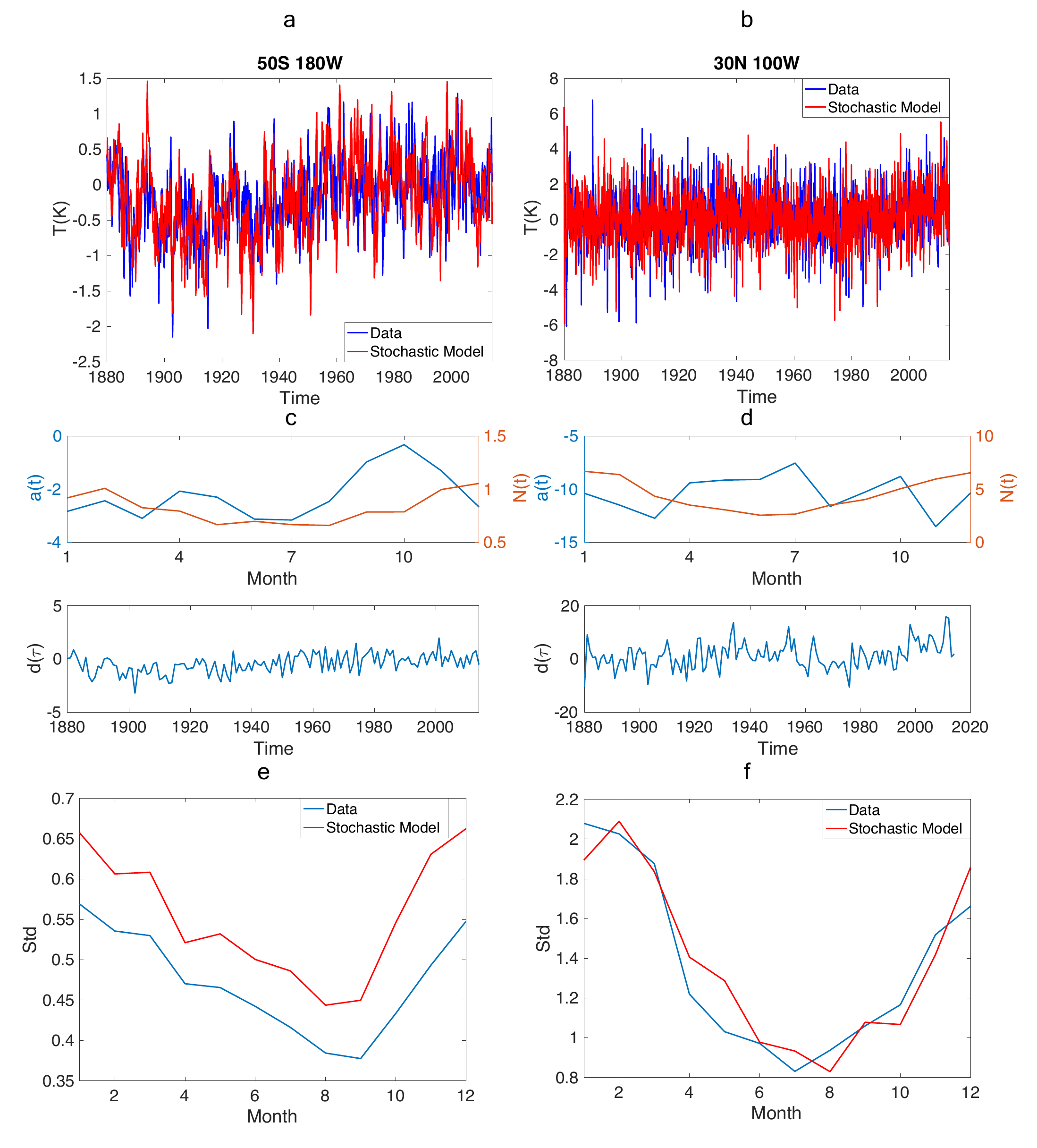} 
\caption{Stochastic models are used to model two monthly time-series chosen from GISS surface temperature data. 
The first (a, c and e) is from one location in the middle of Southern Ocean and the second (b, d and f) is from 
a location in North America. The comparisons between the time-series (blue) and the stochastic model simulation (red)
for the two cases are shown in a and b, respectively. The three time-functions of the stochastic model, $a(t)$ (blue), $N(t)$ (orange) and $d(\tau)$ (blue), are shown in c and d for the two cases. The comparison between the monthly standard deviation for the data (blue) and the model (red) are shown in e and f, respectively.}
\label{fig:sfig01}
\end{figure}

The method is very useful in understanding seasonal variability through the relationship between monthly stability
$a(t)$ and noise intensity $N(t)$.  Figure \ref{fig:sfig01} shows an example of the application of the method to monthly averaged data. We choose two examples of time-series from the GISS surface temperature data; the Southern Ocean (a, c and e) and the southern United States of America (b, d and f). Using the method described above, we 
constructed the two periodic functions $a(t)$ and $N(t)$ and the long-term forcing $d(\tau)$, which are shown for the two cases 
in c and d. Monte Carlo simulations based on the stochastic model $\dot{x}=a(t)x+N(t)\xi+d(\tau)$ are compared with the original time-series in a and b. The comparison between the data and the model simulation is also shown in monthly standard deviation (e and f). It is clear that the model successfully captures the seasonal variability in the data viz.,  $a(t)x+N(t)\xi$ and the long-term variability is associated with the forcing $d(\tau)$. 

\subsection{EOF analysis}

It is shown in the main text that the $1^{st}$ EOF mode constructed from the $d(\tau)$ of the GISS Surface Air Temperature data has pink noise characteristics beyond decadal. In this section, the other major EOF modes (from mode 2 to mode 4) are introduced to confirm that the dominant role of the first mode.    

\begin{figure}
\centering
   \includegraphics[width = 0.45\textwidth]{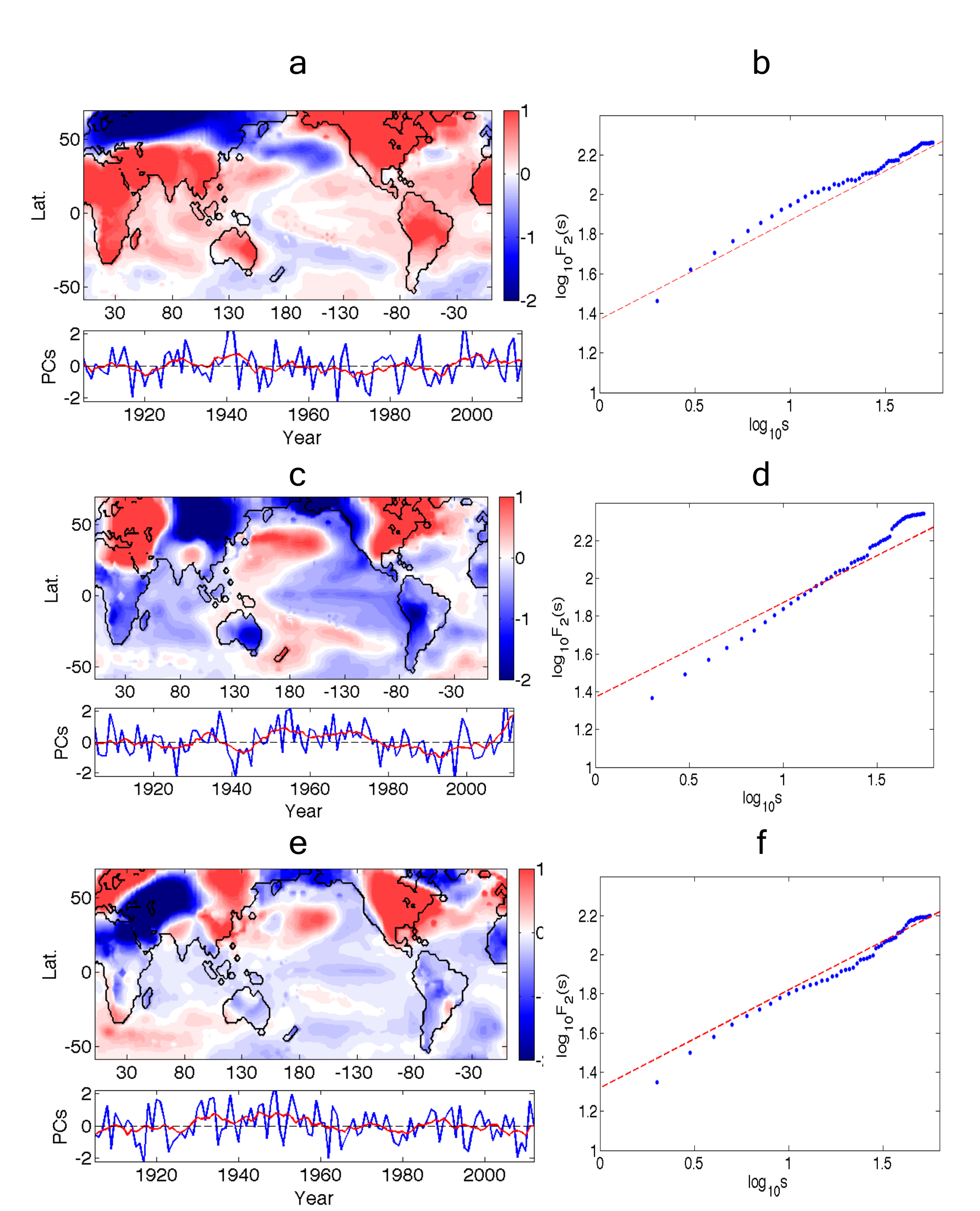}
\caption{The $2^{nd}$ EOF mode of  $d(\tau)$ in the GISS Temperature data, which explains 11.3\% of the total variance. The spatial pattern and the Principal Component series are shown in (a) and the results from MF-TW-DFA are shown in (b), where the red line has a slope of 1/2, which denotes white noise characteristics. (c,d) The $3^{rd}$ EOF mode, which explains 7.4 \% of the total variance. (e,f) The $4^{th}$ EOF mode, which explains 6.5 \% of the total variance.}
\label{fig:EOF}
\end{figure}

The $2^{nd}$ EOF mode constructed from the $d(\tau)$ is shown in  Figure \ref{fig:EOF}(a) to explain the spatial and temporal characteristics of the mode. This mode explains 11.3 \% of the total variance. The spatial pattern can be summarized as the interaction of the PDO in central Pacific Ocean and the low-frequency (wavenumber 1) mode in Northern High Latitudes. The fluctuation function from MF-TW-DFA is shown in  Figure \ref{fig:EOF}(b), exhibiting white noise characteristics beyond decadal (i.e., the fluctuation function parallels the red dotted line on decadal and multi-decadal 
time-scales implying white noise). 

The $3^{rd}$ EOF mode, shown in  Figure \ref{fig:EOF}(c), mirrors the spatial and temporal behavior of the $2^{nd}$ mode. Moreover, it captures the interaction of the PDO in pacific and the low-frequency (wavenumber 1 or 2) in Northern High Latitudes. The positive value in eastern North America and the negative value in west coast North America mirrors the 
Pacific/North American teleconnection mode (PNA). This mode explains around 7\% of total variance. As was the case for the $2^{nd}$ mode, the long-term fluctuation follows white noise characteristics, as is confirmed by the fluctuation function from the MF-TWDFA shown in  Figure \ref{fig:EOF}(d).

The $4^{th}$ EOF mode, which explains about 6\% of total variance is somewhat different from the lower order modes. Principally, this mode exhibits low-frequency (wavenumber 3) structure in the high-latitudes in the Northern hemisphere, but its relationship with the PDO in Central Pacific is not evident. Because this mode can be ascribed to atmospheric motions, one would not expect it to exhibit strong decadal and multi-decadal variability. The whiteness in the decadal and multi-decadal time-scales is clearly shown in the fluctuation function  Figure \ref{fig:EOF}.

Clearly, all of the leading EOF modes except the first exhibit rather short time-scales. Thus, the other modes introduced here cannot contribute to decadal and multi-decadal variability. Therefore, the pink noise characteristics found in various globally distributed proxy data on multi-decadal time-scales should be reflecting the EOF $1^{st}$ mode constructed from the GISS surface air temperature data. The global heat distribution controlled by the major oceanic heat sink regions, such as the Eastern Tropical Pacific and the Southern Ocean provide  the main ingredients in shape the critical behavior seen here on multi-decadal time-scales. 

\subsection{Decadal climate variability reflected in $d(\tau)$}

Working with the quantity $d(\tau)$ has several advantages over using annually-averaged surface air temperature.  

\begin{itemize}

\item Firstly, because it emerges from our periodic non-autonomous stochastic model, it has the interpretation as a forcing term proportional to the excess surface heat flux and thus analysing its temporal and spatial variability is more advantageous than the temperature itself.  

\item Secondly, the stochastic differential equation framework has allowed us to determine it analytically, as shown in Eq. \ref{eq:d}.  

\item Thirdly, a practical advantage resides in addressing the land-ocean contrasts in climate dynamics.  A canonical example is how ENSO affects the weather and
climate of Eurasia or North America. To address this, we first consider the covariance of surface air temperature and observe how the two regions 
are statistically connected. However, the large land/ocean contrast in temperature fluctuations makes this a challenging exercise.
Indeed, even after annually-averaging the temperature, the effect of the land/ocean contrast on seasonal time-scales remains.
Although to examine the impact of ENSO on decadal and multi-decadal time-scales it is desirable to minimize the impact of seasonal fluctuations, annual-averaging 
is not sufficient. In contrast, the quantity $d(\tau)$ is the residual from having filtered out the seasonal fluctuations captured by $a(t)x+N(t)\xi$. Therefore, 
$d(\tau)$ is naturally influenced far less by seasonal variability.

%

For example, assume that the temperature fluctuations in an ENSO region satisfy
$\dot{x}_1=-\lambda_1x_1+N_1\xi_1+d_1(\tau)$. There are two land regions, 2 and 3, satisfying $\dot{x}_{2,3}=-\lambda_{2,3}x_{2,3}+N_{2,3}\xi_{2,3}+d_{2,3}(\tau)$. For simplicity if we assume $\langle \xi_i \xi_j\rangle = \delta_{ij}$, then the covariances between the 
ENSO region and the land regions are cov$( x_1, x_{2,3})\equiv\langle x_1, x_{2,3}\rangle = \frac{1}{\lambda_1\lambda_{2,3}}\langle d_1,d_{2,3}\rangle$, where we see how the stability $\lambda$ controls the covariance. Namely, even if $\langle d_1, d_2\rangle > \langle d_1, d_3\rangle$, it is possible that $\langle x_1, x_2\rangle < \langle x_1, x_3 \rangle$, because of the relative influences of $\lambda_2$ and $\lambda_3$, which are local physical parameters controlling the fluctuations in the local temperature.  Whence, the temperature covariance underlies the influence of ENSO on the local temperature.  However, if our principal goal is to understand the spatial distribution of energy fluxes (e.g., those fluxes associated with ocean upwelling in the Eastern Pacific and Southern Oceans) we are not interested in the local influences on the statistics of the temperature, which makes $d(\tau)$ the appropriate parameter.

\begin{figure}
\centering
   \includegraphics[width = 0.45\textwidth]{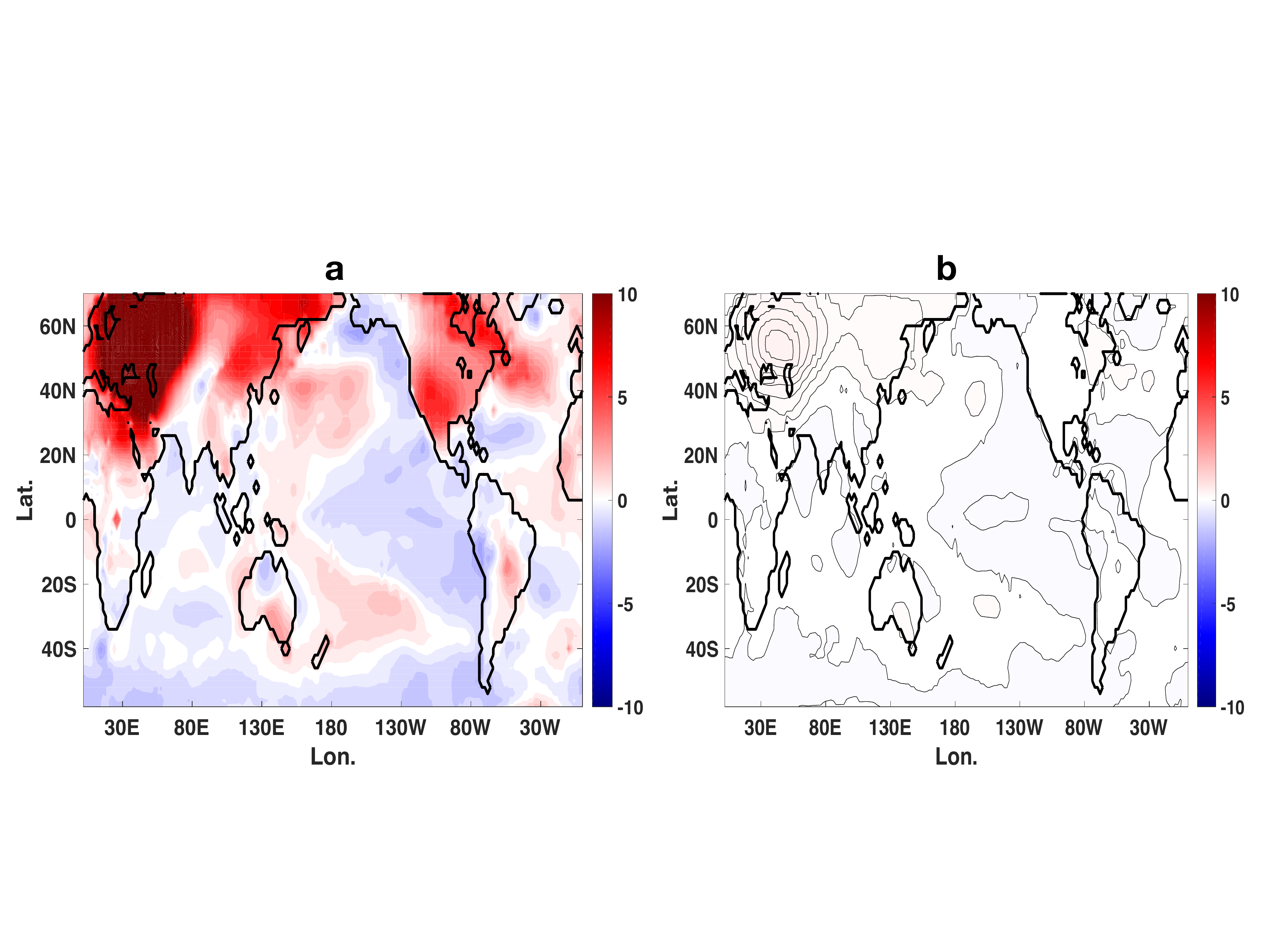} 
\caption{One point covariance global map based on (a) $d(\tau)$ and (b) the annual averaged temperature $\bar{T}$. 
 The center point for the covariance map is located in the central Europe. }
\label{fig:cov}
\end{figure}

Figure \ref{fig:cov}(a) shows one-point covariance map of $d(\tau)$ and Figure \ref{fig:cov}(b) the annual average
temperature. These are clearly distinct.  Firstly, the magnitude of the covariance from $d(\tau)$ is significantly larger.  This is because of 
the specific way in which $d(\tau)$ filters out the seasonality discussed above.  Moreover, we observe 
the teleconnection link between Europe and the IPO (or the PDO in Northern Hemisphere) and between Europe and NAO. This is crucial when one
constructs a global spatial mode from EOF analysis, in which the eigenvalues and their eigenmodes are calculated from the covariance matrix, whose 
elements should obviously not be too small in magnitude.  
Therefore, whereas it is possible to find a meaningful global mode from the EOF analysis of $d(\tau)$, it is not possible with the annual average surface air temperature.  
With further emphasize this point in the context of the history of EOF analysis of decadal and multi-decadal climate variability, wherein ostensibly {\em only} Sea Surface Temperature (SST) data has been used.  Thus, the neglect of land masses in such analyses makes it a serious challenge to determine the relationships between oceanic decadal modes (PDO, IPO, or AMO) and the climate of continental regions solely from the EOF principle modes.   In contrast, in our method using $d(\tau)$ does not exclude land regions,
allowing one to see how the decadal or multi-decadal ocean variability affects continental regions. 

\begin{figure}
\centering
   \includegraphics[width = 0.45\textwidth]{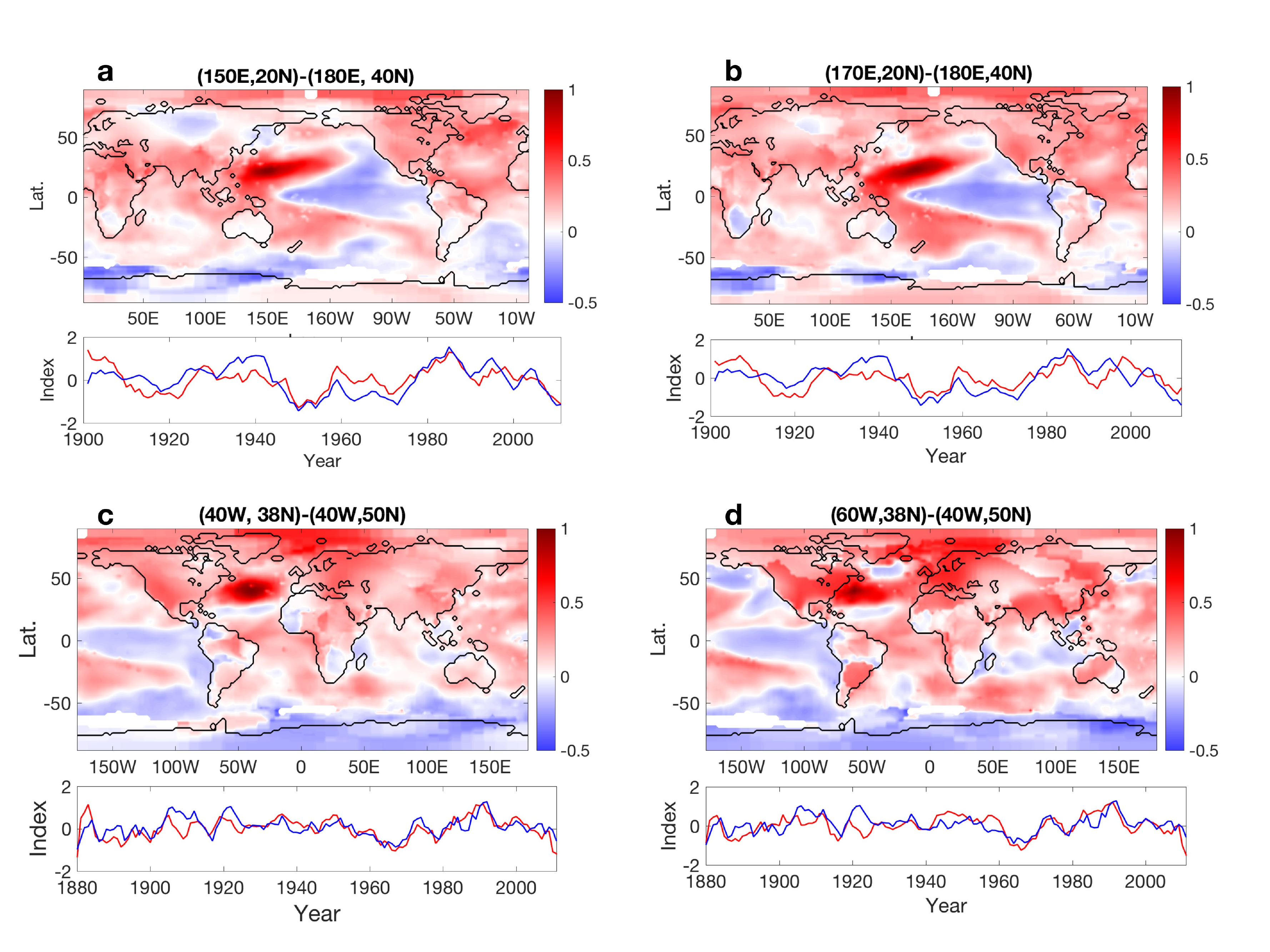} 
\caption{Reconstruction of the indexes of decadal ocean variabilities (PDO, NAO) based on the two-point difference of the $d(\tau)$. Pacific Decadal 
Oscillation index (PDO) is constructed by the two-point difference of $d(\tau)$, which are shown in (a) and (b). The titles represent the locations 
chosen for the two points. In the same way, North Atlantic Oscillation (NAO) indexes are constructed, which are shown in (c) and (d). The constructed 
indexes (red curves) are compared with the original PDO and NAO indexes (blue curves) defined by the PCs of the first EOF mode of the SST in specified ocean regions.
This confirms that there is no sensitivity on choosing two locations for the indexes.}
\label{fig:index}
\end{figure}

\item Another use of $d(\tau)$ concerns the determination of of climate indices representing the decadal and multi-decadal variability of large-scale ocean
circulations. As shown in the one-point covariance map, even though we chose a center point in Europe, we clearly observe the typical
patterns of the PDO and the NAO.  Because $d(\tau)$ filters most of the local seasonality it also acts as a spatial filter, filtering out synoptic and smaller scales and in Fig. 2 of the main text we show that the $d(\tau)$ could replace the original PDO and NAO indices and Figure \ref{fig:index} here shows the robustness of this approach. 
Even though the locations of the two $d(\tau)$'s used for the indices for the PDO (a,b) and the NAO (c,d) differ by several thousands of
kilometers, the new indices, defined by the difference between the two $d(\tau)$'s are robust, showing a reasonable match with the original indices.
By this logic, one can argue that the $d(\tau)$ could replace the original decadal indices, which are constructed based upon EOF analysis. Moreover, 
$d(\tau)$ is directly related to energy flux.     

\end{itemize} 

\subsection{Interpretating scaling properties of interglacial and glacial climates}

\begin{figure}
\centering
   \includegraphics[width = 0.45\textwidth]{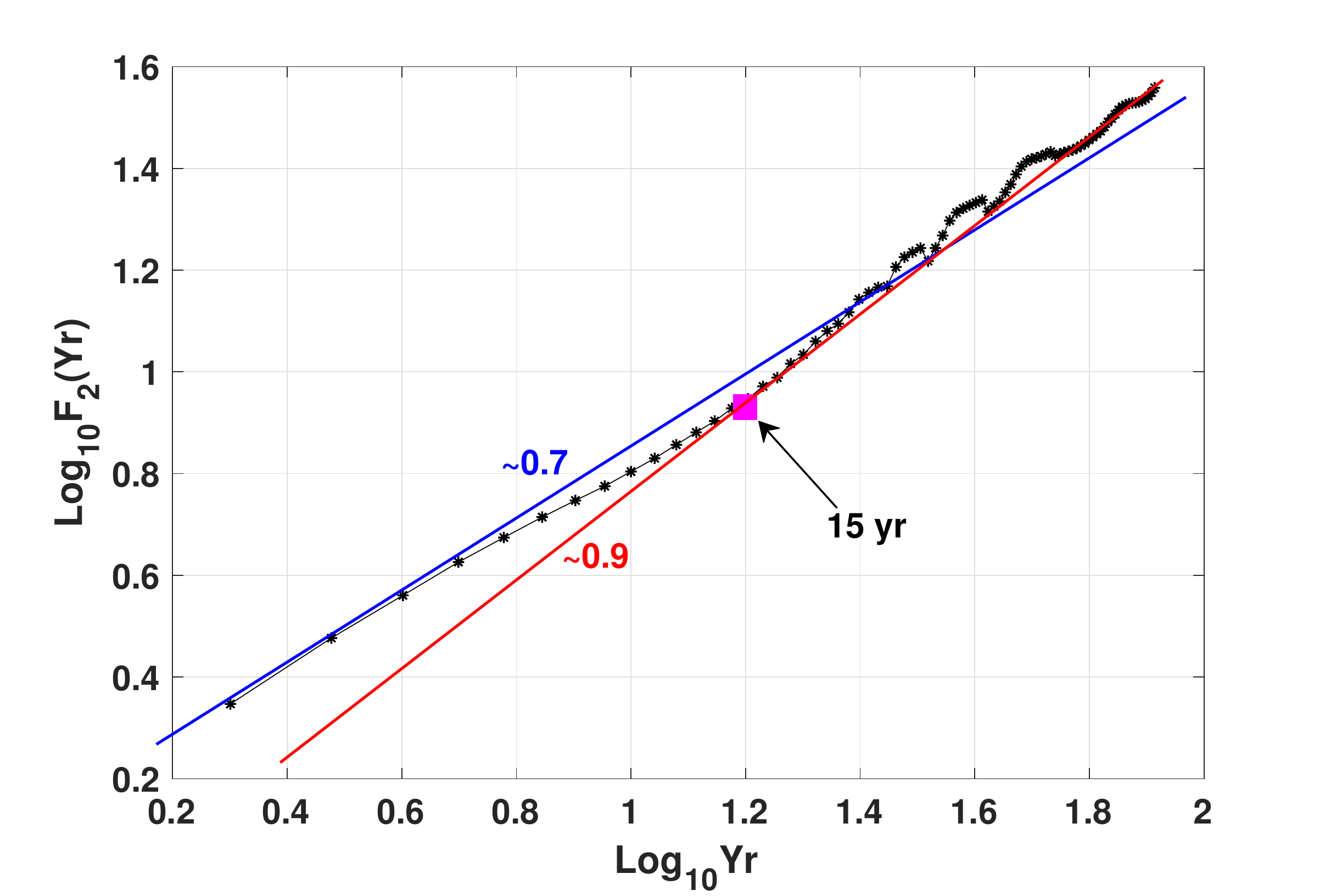} 
\caption{MF-TWDFA analysis using data from a European meteorological station to determine the exponent $\beta$ in $F_2(s) \sim s^{-\beta}$
The blue line has $\beta = 0.7$, deduced using the entire range of data. The red line has $\beta =$ 0.9, consistent with pink noise, and calculated with the data on time-scales beyond the crossover time of 15 years.}
\label{fig:oxford}
\end{figure}

In their analysis of the Holocene record from a European meteorological station \citet{shao2016} determine the exponent $\beta=0.7$ in $F_2(s) \sim s^{-\beta}$ over the entire time-range of the data.  On the other hand, we interpret that 
there is a crossover located $\sim$ 15-years, after which the slope changes to $\beta =$ 0.9, consistent with pink noise.  This comparison is shown in Figure \ref{fig:oxford}, where the blue line has $\beta = 0.7$ and the red line has slope $\beta = 0.9$.

%


\begin{table*}
\fontsize{7}{7}\selectfont
\centering
\caption{Description of data used in this study}
\label{paleotable}
\begin{tabular}{lll}
\hline
Region & Ice Core / Speleothem & Dataset                                                                                                          \\ \hline
Antarctica      & Ice Core                       & \begin{tabular}[c]{@{}l@{}}James Ross Island Ice Core 14000 Year Deuterium Temperature Data\end{tabular}               \\ 
Antarctica      & Ice Core                       & Antarctic Peninsula Gomez Ice Core 150 Year d18O Data                                                                     \\ 
Antarctica      & Ice Core                       & \begin{tabular}[c]{@{}l@{}}Antarctic Ice Core Deglacial Water Isotope Data on GICC05 Timescale\end{tabular}            \\ 
Antarctica      & Ice Core                       & EPICA Dome C Ice Core nss-Ca and Na Data                                                                                  \\ 
Antarctica      & Ice Core                       & \begin{tabular}[c]{@{}l@{}}EPICA Dome C Ice Core 800KYr Dust Flux Data at 25yr Resolution\end{tabular}                 \\ 
Antarctica      & Ice Core                       & \begin{tabular}[c]{@{}l@{}}Mt. Erebus Saddle, Antarctica 500 Year Ice Core Glaciochemical Data\end{tabular}            \\ 
Antarctica      & Ice Core                       & Law Dome Ice Core 2000-Year CO2, CH4, and N2O Data                                                                        \\ 
Antarctica      & Ice Core                       & TALDICE Ice Core 150KYr 1cm Calcium and Sodium Data                                                                       \\ 
Antarctica      & Ice Core                       & Talos Dome Ice Core Deuterium Data                                                                                        \\ 
Antarctica      & Ice Core                       & \begin{tabular}[c]{@{}l@{}}Dome C 2500 Year VOLSOL Ice Core Sulfate Concentration Data\end{tabular}                    \\ 
Antarctica      & Ice Core                       & WAIS Divide Ice Core 67.2-9.8 kaBP Continuous CH4 Data                                                                    \\ 
Asia            & Ice Core                       & Dasuopu, China Ice Core Oxygen Isotope Data                                                                               \\ 
Asia            & Ice Core                       & Dunde, China Ice Core Oxygen Isotope Data                                                                                 \\ 
Asia            & Ice Core                       & \begin{tabular}[c]{@{}l@{}}East Rongbuk Glacier Col Ice Core 1,000 Year Geochemical Data\end{tabular}                  \\ 
Asia            & Speleothem                     & \begin{tabular}[c]{@{}l@{}}2650-Year Beijing Stalagmite Layer Thickness and Temperature Reconstruction\end{tabular}    \\ 
Asia            & Speleothem                     & \begin{tabular}[c]{@{}l@{}}Central and Northeast India 1000 Year Stalagmite Oxygen Isotope Data\end{tabular}           \\ 
Asia            & Ice Core                       & \begin{tabular}[c]{@{}l@{}}Dasuopu Ice Core 1000 Year d18O, Dust, Anion and Accumulation Data\end{tabular}             \\ 
North America   & Ice Core                       & Camp Century Ice Core Data                                                                                                \\ 
North America   & Ice Core                       & Crete Ice Core Data                                                                                                       \\ 
North America   & Ice Core                       & Agassiz Delta 18-O Annuals                                                                                                \\ 
North America   & Ice Core                       & Agassiz Delta 18-O 25 Year Averages                                                                                       \\ 
North America   & Ice Core                       & Agassiz 1984 Delta 18-O Annual Averages                                                                                   \\ 
North America   & Ice Core                       & Agassiz 1984 Ecm 10 Year Averages                                                                                         \\ 
North America   & Ice Core                       & GISP2 Oxygen Isotope Data (1 year averages)                                                                               \\ 
North America   & Ice Core                       & GISP2 Bidecadal Oxygen Isotope Data                                                                                       \\ 
North America   & Ice Core                       & DYE 3 DELTA 18-O ANNUAL AVERAGES                                                                                          \\ 
North America   & Ice Core                       & \begin{tabular}[c]{@{}l@{}}GISP2 Ice Core 4000 Year Ar-N Isotope Temperature Reconstruction\end{tabular}  \\ 
North America   & Ice Core                       & NEEM-2011-S1 Ice Core 1800 Year Continuous Methane Data                                                                   \\ 
North America   & Ice Core                       & Combined NGRIP-Cariaco Basin 56,000 Year d18O Record                                                                      \\ 
North America   & Ice Core                       & \begin{tabular}[c]{@{}l@{}}Renland and Agassiz Ice Core 60KYr d18O Data on GICC05 Timescale\end{tabular}               \\ 
North America   & Speleothem                     & \begin{tabular}[c]{@{}l@{}}Crystal Cave, California 1150 Year d18O Data and SST Reconstruction\end{tabular}            \\ 
Central America & Speleothem                     & \begin{tabular}[c]{@{}l@{}}Yok Balum Cave, Belize 2000 Year Stalagmite Stable Isotope Data\end{tabular}                \\ 
Central America & Speleothem                     & \begin{tabular}[c]{@{}l@{}}Juxtlahuaca Cave, Mexico 2400 Year d18O and Precipitation Reconstruction\end{tabular}       \\ 
South America   & Ice Core                       & \begin{tabular}[c]{@{}l@{}}Quelccaya Ice Cap, Peru 1800 Year Oxygen Isotope, Dust, and Major Ions Data\end{tabular}    \\ 
Europe          & Speleothem                     & \begin{tabular}[c]{@{}l@{}}Spannagel Cave Oxygen Isotope (d18O) Data for the last 10.8ka\end{tabular}                  \\ 
Europe          & Speleothem                     & \begin{tabular}[c]{@{}l@{}}Klapferloch Cave, Austria 3000 Year Speleothem Stable Isotope Data\end{tabular}             \\ 
Europe          & Speleothem                     & \begin{tabular}[c]{@{}l@{}}Northwest Scotland Stalagmite and Climate Reconstruction Data\end{tabular}                  \\ 
Europe          & Speleothem                     & \begin{tabular}[c]{@{}l@{}}Sieben Hengste Cave, Switzerland LGM Stalagmite Oxygen Isotope Data\end{tabular}            \\ 
Europe          & Speleothem                     & \begin{tabular}[c]{@{}l@{}}Uamh an Tartair Cave, Northwest Scotland 3000 Year Speleothem Growth Rate Data\end{tabular} \\ 
Europe          & Ice Core                       & Svalbard Ice Cores 600 Year Annual d18O Data                                                                              \\ 
Pacific         & Speleothem                     & \begin{tabular}[c]{@{}l@{}}Espiritu Santo, Vanuatu 446 Year Stalagmite Oxygen Isotope Data\end{tabular}                \\ 
Pacific         & Speleothem                     & Guadalcanal Speleothem 600 Year Stable Isotope Data                                                                       \\ \hline
\end{tabular}
\end{table*}

\begin{figure*}
\centering
   \includegraphics[width = 0.8\textwidth]{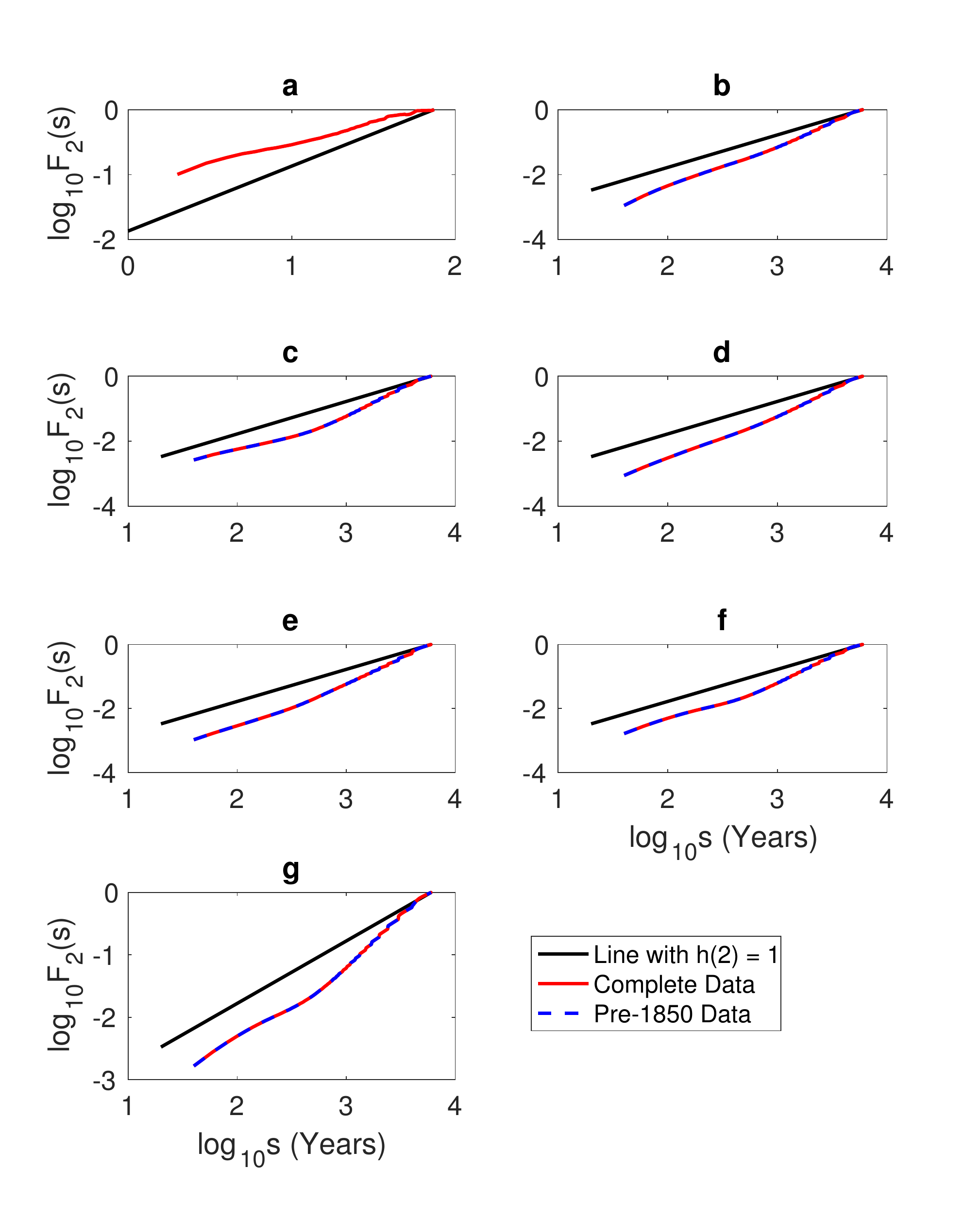}
\caption{Fluctuation functions for the $\delta ^{18}O$ isotope in Antarctica for (a) the complete record from the Gomez Ice Core from 1857 - 2005 AD, (b) Byrd Dome from 32 - 8 kyr BP (1950), (c) EPICA Dronning Maud Land  core from 32 - 8 kyr BP (1950) , (d) Law Dome from 32 - 8 kyr BP (1950) , (e) Siple Dome from 32 - 8 kyr BP (1950) , (f) Talos Dome from 32 - 8 kyr BP (1950), and (g) Mt. Erebus Saddle from 1472 - 2006 AD.}
\label{fig:o18A}
\end{figure*}

\begin{figure*}
\centering
	\includegraphics[width = 0.8\textwidth]{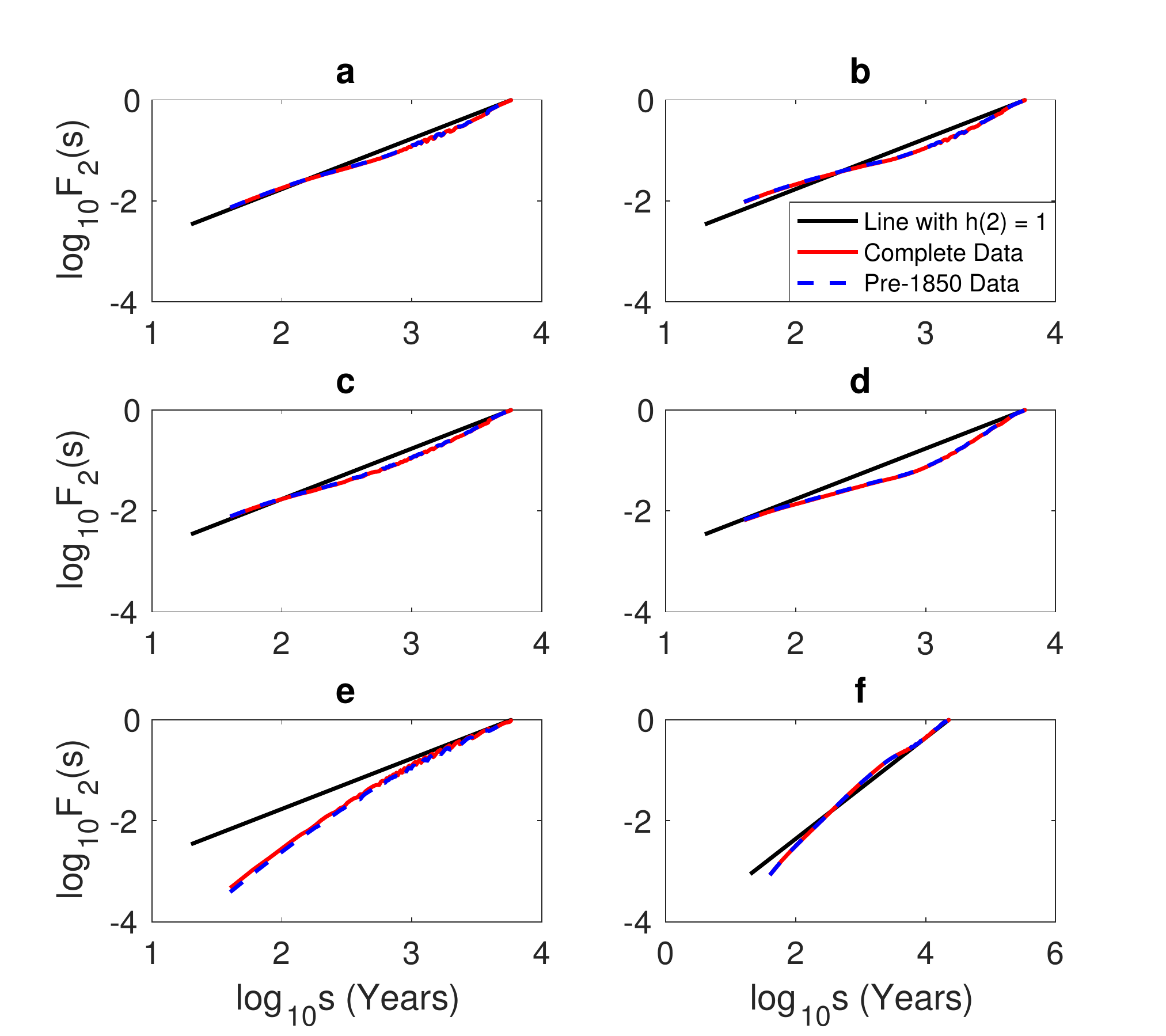}
\caption{Fluctuation functions for the $\delta ^{18}O$ isotope for the complete and Pre-1850 records in Greenland from 0 - 11.7ka before 2000 AD with 20 years resolution for (a) Agassiz 1987 ice core, (b) Agassiz 1984 ice core, (c) Agassiz 1979 ice core, (d) Agassiz 1977 ice core, (e) Renland ice core from 0 - 59.44ka before 2000 AD, and (f) Combined NGRIP-Cariaco Basin from 10650 - 56050 years before present}
\label{fig:o18G1}
\end{figure*}

\begin{figure*}
\centering
   \includegraphics[width = 0.8\textwidth] {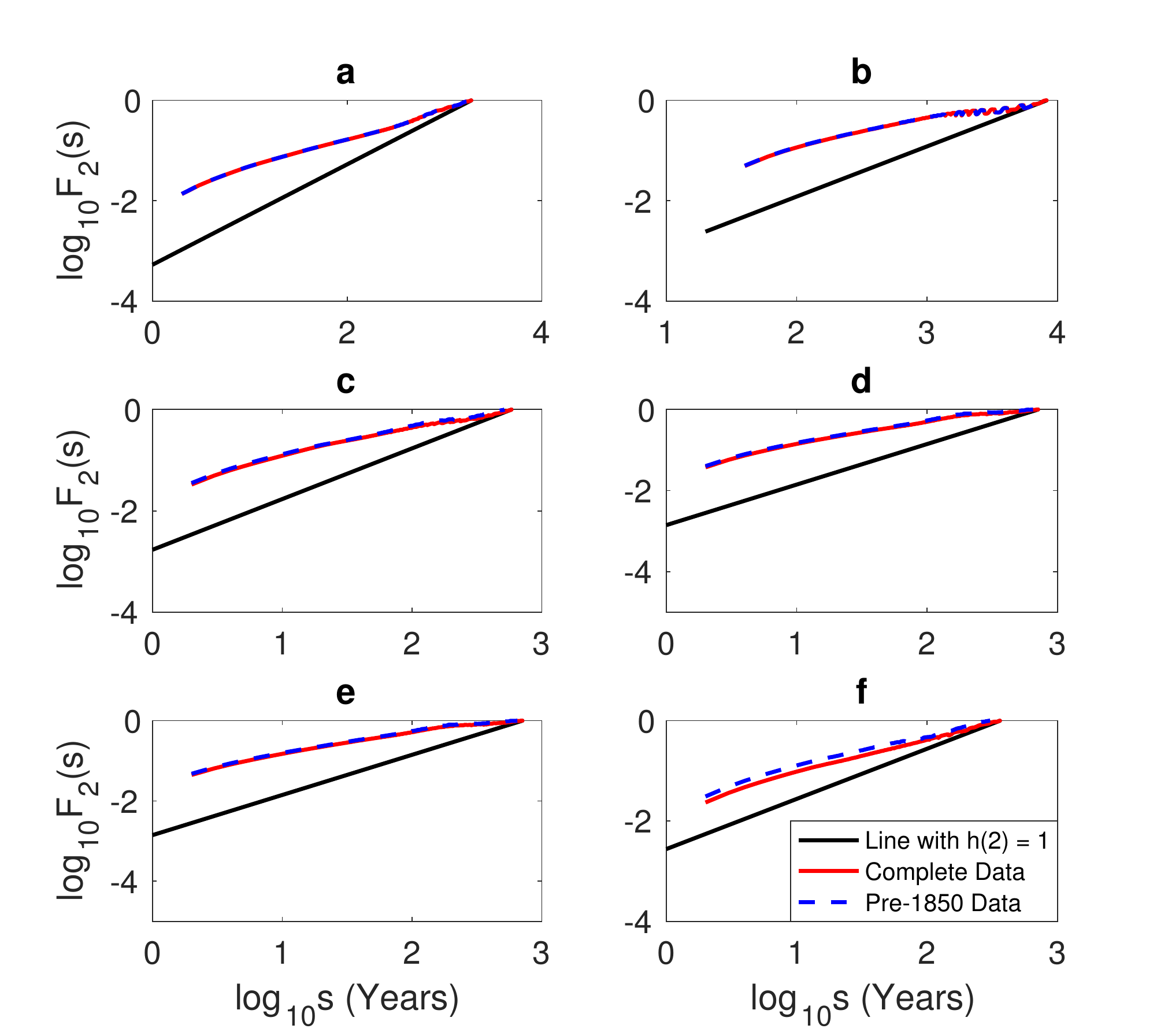}
\caption{Fluctuation functions for the $\delta ^{18}O$ isotope for the complete and Pre-1850 records in Greenland for (a) DYE 3 ice core from 1899BC - 1872AD, (b) GISP2 from -30 - 16510 years before 1950, (c) GISP2 from -37 - 1133 years before 1950, (d) Crete 1974 ice core from 553 - 1974 AD, (e) Crete 1974 ice core from 553 - 1974 AD with 12 samples per year, and (f) Camp Century ice core from 1242 - 1967 AD.}
\label{fig:o18G2}
\end{figure*}

\begin{figure*}
\centering
   \includegraphics[width = 0.8\textwidth]{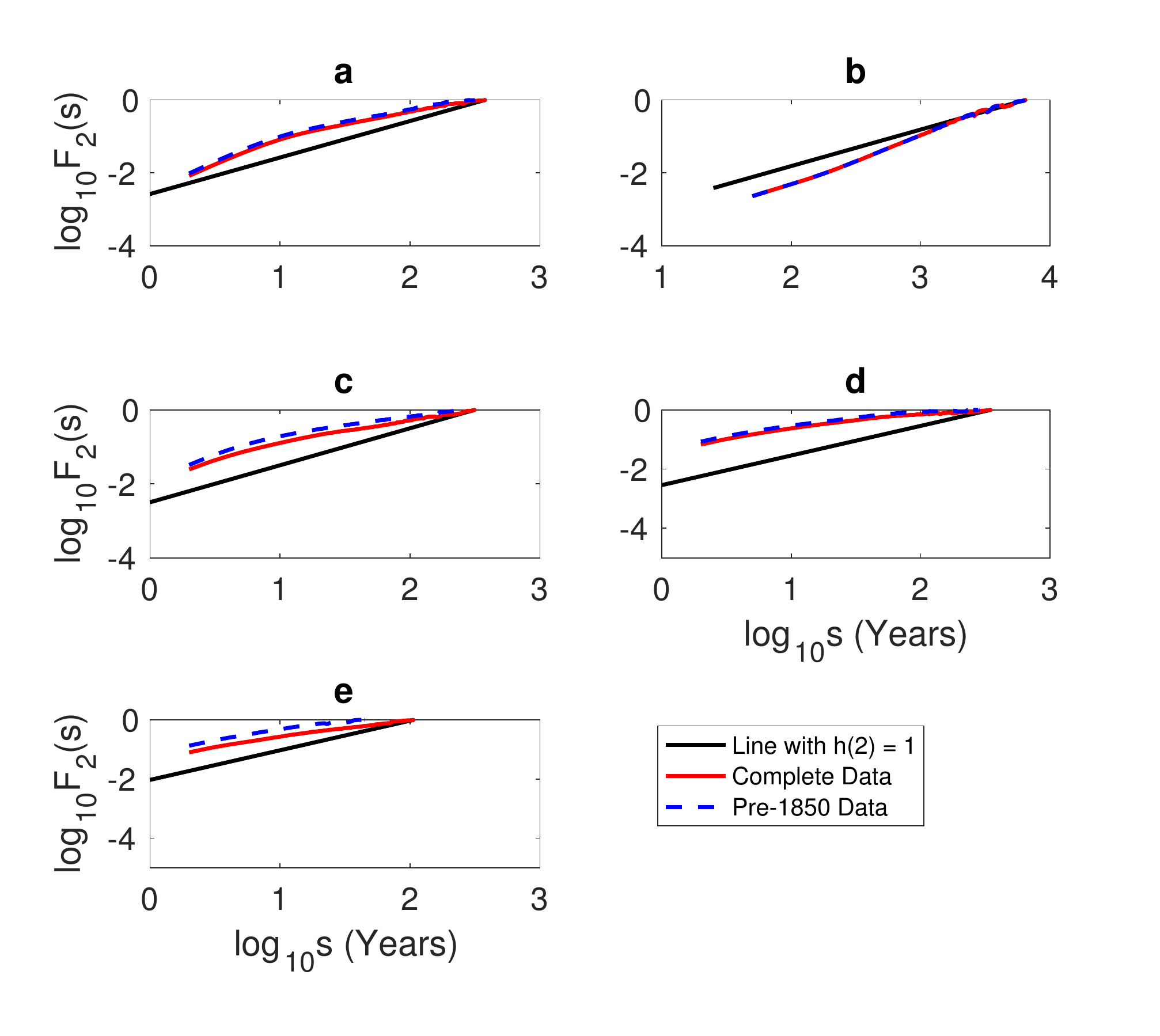}
\caption{Fluctuation functions for the $\delta ^{18}O$ isotope for the complete and Pre-1850 records in Greenland for (a) Agassiz 1984 ice core from 1217 - 1973 AD, (b) Agassiz 1977 ice core from 10973 BC - 1977 AD, (c) Agassiz 1977 ice core from 1349 - 1977 AD, (d) Crete 1974 ice core from 1278 - 1974 AD with 12 samples per year, and (e) Camp Century ice core with annual resolution from 1761 - 1975 AD.}
\label{fig:o18G3}
\end{figure*}

\end{document}